\documentclass[10pt,journal,compsoc]{IEEEtran}
	
\usepackage{amsmath,amsfonts}
\usepackage{algorithmic}
\usepackage{graphicx}
\usepackage{textcomp}
\usepackage{xcolor}

\usepackage{etex}
\usepackage{booktabs}
\usepackage[utf8]{inputenc}
\usepackage[ruled,vlined]{algorithm2e}
\usepackage[T1]{fontenc}
\usepackage{microtype}
\usepackage{graphicx}
\usepackage{graphics}
\usepackage{paralist}
\usepackage{tabularx}
\usepackage{balance}
\usepackage{multicol}
\usepackage{multirow}
\usepackage{pbox}
\usepackage{enumitem}
\usepackage{lscape}
\usepackage{pifont}
\usepackage{xspace}
\usepackage{url}
\usepackage{tikz}
\usepackage{float}
\usepackage[TABBOTCAP]{subfigure}
\usepackage{ragged2e}
\usepackage{fontawesome}
\usepackage[figuresright]{rotating}
\usepackage{bbding} 

\usepackage{adjustbox}

\ifCLASSOPTIONcompsoc
  \usepackage[nocompress]{cite}
\else
  \usepackage{cite}
\fi

  {\list{}{\leftmargin=0.1in\rightmargin=0.1in}\item[]}%
  {\endlist}

\newboolean{showcomments}

\setboolean{showcomments}{true}

\ifthenelse{\boolean{showcomments}}
  {\newcommand{\nb}[2]{
    \fbox{\bfseries\sffamily\scriptsize#1}
    {\sf\small$\blacktriangleright$\textit{#2}$\blacktriangleleft$}
   }
  }
  {\newcommand{\nb}[2]{}
  }

\newcommand{\ie}{\emph{i.e.,}\xspace}
\newcommand{\eg}{\emph{e.g.,}\xspace}
\newcommand{\etc}{etc.\xspace}
\newcommand{\etal}{\emph{et~al.}\xspace}
\newcommand{\secref}[1]{Section~\ref{#1}\xspace}

\newcommand{\figref}[1]{Fig.~\ref{#1}\xspace}

\newcommand{\tabref}[1]{Table~\ref{#1}\xspace}

\newcommand{\manually}{2,291\xspace}
\newcommand{\conflicts}{1,225\xspace}
\newcommand{\categories}{120\xspace}
\newcommand{\labels}{120\xspace}
\newcommand{\cc}{\emph{code \& comment-to-code}\xspace}
\newcommand{\cnl}{\emph{code-to-comment}\xspace}

\newcommand{\tufano}{{\scshape T5cr}\xspace}
\newcommand{\li}{{\scshape CodeReviewer}\xspace}
\newcommand{\hong}{{\scshape CommentFinder}\xspace}

\newcommand{\rev}[1]{\textcolor{black}{#1}}
\newcommand{\revs}[1]{\textcolor{black}{#1}}

\begin{document}

\title{Code Review Automation: Strengths and Weaknesses of the State of the Art}

\author{Rosalia~Tufano,
	Ozren~Dabi\'c,
	Antonio~Mastropaolo,
	Matteo~Ciniselli,
	and~Gabriele~Bavota
	\IEEEcompsocitemizethanks{\IEEEcompsocthanksitem R. Tufano is with SEART @  Software Institute, Universit\`a della Svizzera italiana, Switzerland. \protect\\
		E-mail: rosalia.tufano@usi.ch
		\IEEEcompsocthanksitem O. Dabi\'c is with SEART @  Software Institute, Universit\`a della Svizzera italiana, Switzerland. \protect\\
		E-mail: ozren.dabic@usi.ch
		\IEEEcompsocthanksitem A. Mastropaolo is with  SEART @  Software Institute, Universit\`a della Svizzera italiana, Switzerland. \protect\\
		E-mail: antonio.mastropaolo@usi.ch
		\IEEEcompsocthanksitem M. Ciniselli is with  SEART @  Software Institute, Universit\`a della Svizzera italiana, Switzerland. \protect\\
		E-mail: matteo.ciniselli@usi.ch
		\IEEEcompsocthanksitem G. Bavota is with SEART @  Software Institute, Universit\`a della Svizzera italiana, Switzerland. \protect\\
	E-mail: gabriele.bavota@usi.ch}
}

\markboth{Journal of \LaTeX\ Class Files,~Vol.~xx, No.~x, Month~xxxx}%
{Tufano \MakeLowercase{\etal}: Code Review Automation:\\Strengths and Weaknesses of the State of the Art}
\IEEEtitleabstractindextext{%
\begin{abstract}
The automation of code review has been tackled by several researchers with the goal of reducing its cost. The adoption of deep learning in software engineering pushed the automation to new boundaries, with techniques \emph{imitating} developers in generative tasks, such as commenting on a code change as a reviewer would do or addressing a reviewer's comment by modifying code. The performance of these techniques is usually assessed through quantitative metrics, \eg the percentage of instances in the test set for which correct predictions are generated, leaving many open questions on the techniques' capabilities. For example, knowing that an approach is able to correctly address a reviewer's comment in 10\% of cases is of little value without knowing what was asked by the reviewer: What if in all successful cases the code change required to address the comment was just the removal of an empty line? In this paper we aim at characterizing the cases in which three code review automation techniques tend to succeed or fail in the two above-described tasks. \rev{The study has a strong qualitative focus, with $\sim$105 man-hours of manual inspection invested in manually analyzing correct and wrong predictions generated by the three techniques, for a total of \manually inspected predictions. The output of this analysis are two taxonomies reporting, for each of the two tasks, the types of code changes on which the experimented techniques tend to succeed or to fail, pointing to areas for future work.} A result of our manual analysis was also the identification of several issues in the datasets used to train and test the experimented techniques. Finally, we assess the importance of researching in techniques specialized for code review automation by comparing their performance with ChatGPT, a general purpose large language model\rev{, finding that ChatGPT struggles in commenting code as a human reviewer would do.}
\end{abstract}

	}

\maketitle

\IEEEdisplaynontitleabstractindextext
\IEEEpeerreviewmaketitle

\section{Introduction} \label{sec:intro}

The benefits of code review are well-known and supported by empirical evidence \cite{bacchelli2013expectations,McIntosh:msr2014,morales2015saner,Bavota:icsme2015,Caitlin:icse2018}. However, works in the literature also documented the non-trivial costs of such a process \cite{Rigby:fse2013,Rigby:tosem2014,Bosu:2013}. To bring down such a cost, researchers proposed techniques to (partially) automate  code review tasks. Originally, most of the approaches aimed at recommending appropriate reviewers for a given code change \cite{al2020workload,jiang:ist2017,jiang:jss2019,thongtanunam:saner2015,ouni:icsme2016,xia:icsm2015,zanjanitse:2016,rahman:icsec2016,asthana:esecfse2019,mirsaeedi:icse2020,yu:ist2016}. With the rise of Deep Learning (DL) in software engineering, researchers started addressing more challenging generative tasks aimed at \emph{imitating} human developers. Two concrete examples are the tasks automated by Tufano \etal \cite{tufano:icse2022}\footnote{We refer to our previous works in the area as Tufano \etal because the set of authors only partially overlaps.} using a Transformer model. The first, named \cc, aims at automatically implementing a revised code ($C_r$) given the code submitted for review ($C_s$) and a reviewer comment ($R_{nl}$). Thus, the input of the approach is represented by a pair $\langle C_s$, $R_{nl} \rangle$, while the output is $C_r$. 

The second, named \cnl, aims at commenting a code submitted for review as a reviewer would do: The input of the approach is $C_s$, while the expected output is $R_{nl}$ (\ie a natural language comment asking for changes to $C_s$). 

The work by Tufano \etal \cite{tufano:icse2022} is only the first of several focusing on automating one or both of these tasks \cite{tufano:icse2022,li_lingwei:esecfse2022,hong:fse2022,li:esecfse2022}. These techniques have been evaluated on test sets containing hundreds of instances representative of the automated tasks. For example, for the \cc task, the test sets feature $\langle C_s$, $R_{nl} \rangle$ pairs which are fed to the approach to assess whether it can address the reviewer's comment $R_{nl}$ and generate the expected $C_r$. The outcome of these evaluations is a mostly quantitative report showing, \eg the percentage of instances in the test set for which the approach successfully generated a prediction. However, such quantitative measures only tell part of the story. Indeed, it could happen that the approach is targeting the \emph{low-hanging fruits}, being successful in only simple code review scenarios which are unlikely to save developers' time. For the \cc task this might mean successfully addressing mostly comments requiring minor changes to $C_s$ (\eg addition/removal of whitespaces to improve the formatting). Similarly, for the \cnl task the approach could overfit and mostly be successful in posting comments related to \eg replacing the \texttt{==} operator in Java with an \texttt{equals} invocation when needed. In other words, little is known about the code review scenarios in which these techniques succeed or fail.  

To fill this gap, we manually analyzed \manually predictions generated by three state-of-the-art techniques \cite{tufano:icse2022,hong:fse2022,li:esecfse2022} automating the code review tasks previously described. The predictions have been generated on the original test sets used in the papers presenting the subject techniques. 

The result of such an analysis are two taxonomies (one per task) featuring a total of \categories types of code changes requested during code review (\eg \emph{extract method refactoring}, \emph{add thrown exception}) with indication about the extent to which state-of-the-art techniques are successful in (i) requesting their implementation when needed by properly commenting the submitted code as a human reviewer would do (\cnl task); (ii) automatically implementing them to address a reviewer's comment (\cc task). 

We found that the proposed techniques can provide support in a wide variety of code changes. However, there are areas of our taxonomies in which the approaches consistently fail, pointing to the need for more research. \rev{As a concrete example, the experimented techniques struggle when they need to recommend (\cnl task) or implement (\cc task) complex code changes spanning across several code components. This is due to the ``view'' they have of the code base, usually limited to a single function or diff hunk submitted for review. This indicates the need for enriching the contextual information provided to these techniques.}

During our manual analysis we also found that $\sim$25\% of the instances in the inspected datasets are the result of data extraction errors possibly undermining the techniques' performance and questioning the validity of the evaluation performed on the test set. We discuss the reasons for such problematic instances.

Finally, given the recent proposal of general purpose Large Language Models (LLMs) such as ChatGPT \cite{chatgpt}, it is unclear what the actual need is for code review automation via specialized techniques. We compared the three subject approaches with ChatGPT, showing that, while the latter represents a competitive solution for the \cc task, it suffers in the \cnl task. 

We release \cite{replication} all data used in and output of our study. 
\section{Related Work} \label{sec:related}

\begin{table*}
\centering
\caption{Summary of related work. For Li Z. \etal \cite{li:esecfse2022} the size of the test set refers to Java instances only.}
\scriptsize
\begin{tabular}{l>{\centering\arraybackslash}p{1.5cm}>{\centering\arraybackslash}p{2.5cm}>{\centering\arraybackslash}p{2.2cm}>{\centering\arraybackslash}p{2.2cm}>{\centering\arraybackslash}p{2cm}>{\centering\arraybackslash}p{1.5cm}} 
\toprule
\textbf{Reference} & \textbf{Venue} & \textbf{Task} & \textbf{Granularity} & \textbf{Underlying technique} & \textbf{\# Training instances} & \textbf{\# Test instances} \\
\midrule
\textbf{Tufano M. \etal \cite{tufano:icse2019}}  & ICSE'19   & \emph{code-to-code} & method   & NMT & 8.6k & 1k\\
\cmidrule{1-7}
\multirow{2}{*}{\textbf{Tufano R. \etal \cite{tufano:icse2021}} } & \multirow{2}{*}{ICSE'21} & \emph{code-to-code}& \multirow{2}{*}{method} & \multirow{2}{*}{NMT} & \multirow{2}{*}{13.7k} & \multirow{2}{*}{1.7k}\\
                                                                   && \emph{code \& comment-to-code} &&&&\\
\cmidrule{1-7}
\multirow{3}{*}{\textbf{Tufano R. \etal \cite{tufano:icse2022}}} & \multirow{3}{*}{ICSE'22} & \emph{code-to-code}  & \multirow{3}{*}{method} & \multirow{3}{*}{T5 (pre-trained)} & \multirow{3}{*}{134.2k} & \multirow{3}{*}{16.7k}\\
                                                                  && \emph{code \& comment-to-code} &&&&\\
                                                                  && \emph{code-to-comment} &&&&\\
\cmidrule{1-7}
\textbf{Thongtanunam \etal \cite{thongtanunam:icse2022}} & ICSE'22         & \emph{code-to-code}  & method & NMT  & 118k & 14.7k\\
\cmidrule{1-7}

\multirow{2}{*}{\textbf{Li L. \etal \cite{li_lingwei:esecfse2022}}} & \multirow{2}{*}{ESEC/FSE'22} & \emph{code \& comment-to-code}  & \multirow{2}{*}{method} & \multirow{2}{*}{T5 (pre-trained)}   & \multirow{2}{*}{87k} & \multirow{2}{*}{1k}\\
                                                                  && \emph{code-to-comment} &&&&\\
\cmidrule{1-7}
\textbf{Hong \etal \cite{hong:fse2022}}                & ESEC/FSE'22 & \emph{code-to-comment}   & method  & IR     & 13.7k & 1.7k\\
\cmidrule{1-7}
\multirow{3}{*}{\textbf{Li Z. \etal \cite{li:esecfse2022}}} & \multirow{3}{*}{ESEC/FSE'22} & \emph{code \& comment-to-code} & \multirow{3}{*}{diff hunk} & \multirow{3}{*}{T5 (pre-trained)} & 117.7k & 2.2k\\
                                                          & & \emph{code-to-comment} &&& 150.4k &1.6k\\
                                                          & & \textit{code quality estimation} &&& 265.8k & 66.4k\\
\bottomrule
\end{tabular}
\label{tab:related}
\end{table*}

Most of the works in code review automation aim at recommending the most suited reviewer for a given change \cite{al2020workload,chouchen:asc2021,jiang:ist2017,jiang:jss2019,strand:icseseip2020,thongtanunam:saner2015,ouni:icsme2016,xia:icsm2015,zanjanitse:2016,rahman:icsec2016,ying:csise2016,xia:iwsm2017,asthana:esecfse2019,mirsaeedi:icse2020,yu:ist2016}. These works differ in the features and algorithms used for the recommendation. Other techniques focus on the binary classification of the quality of the code submitted for review (\ie whether the change should be accepted or not) using machine \cite{chouchen:asc2021} or deep \cite{shu:aaai2019,li:esecfse2022} learning. Our work is however mostly related to the techniques aimed at \emph{imitating} human reviewers by automatically reviewing the submitted code. Thus, we focus our discussion on these techniques which are summarized in \tabref{tab:related}. 

Tufano M. \etal \cite{tufano:icse2019} proposed the usage of Neural Machine Translation (NMT) to learn how to automatically modify a given Java method as developers would do during a pull request (PR): The NMT model takes as input a method submitted in a PR and implements code changes likely to be required during the review of the PR.

Tufano R. \etal \cite{tufano:icse2021} built on top of this idea by presenting two transformer-based models to automate two code review tasks. The first, named \emph{code-to-code}, is a replica of the task automated by Tufano M. \etal \cite{tufano:icse2019}: It takes as input a Java method submitted for review ($C_s$) and implements changes likely to be required during code review, producing a revised method $C_r$. The second is the already discussed \cc task. 

Both of the aforementioned works suffered of a major limitation: They relied on a code abstraction process to reduce the vocabulary size and simplify the learning of the DL models. This means that the models did not work on raw code, but on an abstracted version of it in which, for example, variable names were replaced with a \texttt{VAR\_ID} token, with \texttt{ID} being a progressive number going from 1 to $n$ ($n$ = number of variables declared in the method). However, such an abstraction had a cost to pay, since the models were not able to implement code changes requiring the introduction of new identifiers and literals. To understand this point let us consider the \emph{code-to-code} task, in which the model takes as input a submitted method $C_s$ and generates a revised method $C_r$. Both the input and the generated method are abstracted. The abstraction process applied to $C_s$ allows to map each identifier to its abstracted version. For example, it is possible to know that a variable \texttt{sum} in $C_s$ has been mapped to \texttt{VAR\_1}. Thus, if the generated $C_r$ method features \texttt{VAR\_1}, it can be mapped back to \texttt{sum}, making $C_r$ understandable by the developer receiving such a recommendation. Differently, if $C_r$ features a new identifier \texttt{VAR\_4} which was not present in $C_s$, it is not possible to map it to raw code, making the recommendation useless. For this reason, both of these works \cite{tufano:icse2019,tufano:icse2021} only automated very simple code review tasks, mostly requiring re-arranging the code tokens in $C_s$ and avoiding the introduction of new identifiers and literals.

To address this limitation, Tufano R. \etal \cite{tufano:icse2022} proposed a new approach which exploited the Text-To-Text Transfer Transformer (T5) model \cite{raffel:jmlr2019} and adopted a SentencePiece tokenizer \cite{Kudo:sentencePiece}, allowing the model to work with raw source code by keeping under control the vocabulary size, avoiding the need for abstraction. Besides reporting better performance as compared to their previous approach \cite{tufano:icse2021}, this work is also the first one attempting the automation of the \cnl task (\ie commenting the code to request changes as a human reviewer would do). 

Thongtanunam \etal \cite{thongtanunam:icse2022} also used transformers to automate code review. They focused on the \emph{code-to-code} task, comparing with the original work by Tufano M. \etal \cite{tufano:icse2019} and showing the superiority of their approach.

Li L. \etal \cite{li_lingwei:esecfse2022} relied on pre-trained DL models to improve the performance achieved by Tufano R. \etal \cite{tufano:icse2022} on the \cc and the \cnl tasks. Hong \etal \cite{hong:fse2022} showed instead that a simpler approach based on IR can achieve competitive performance as compared to DL models when it comes to the \cnl task: Rather than generating a reviewer's comment for a submitted Java method (as done by the DL-based techniques), the approach by Hong \etal \cite{hong:fse2022} identifies in the ``training set'' the method $C_{ms}$ being most similar to the $C_s$ submitted for review. Based on this information, it retrieves review comments which have been posted in the past for $C_{ms}$, assuming they would be valuable for $C_s$ as well. Both Li L. \etal \cite{li_lingwei:esecfse2022} and Hong \etal \cite{hong:fse2022} show the superiority of their technique as compared to the approach by Tufano R. \etal \cite{tufano:icse2022}.

Finally, Li Z. \etal  \cite{li:esecfse2022} targeted the automation of three code review tasks. The first among these represent the binary classification of submitted code to decide whether it needs a review or can be accepted as is. The other two tasks are \cnl and \cc. One aspect makes this work particularly novel as compared to those previously discussed: While all previous techniques work at method-level granularity (\ie they take as input a method submitted for review), the approach by Li Z. \etal  \cite{li:esecfse2022} takes a ``diff hunk'', namely an area of a specific file modified in a code change to be reviewed. This allows the model to generate file-level ``reviewers' comments'' such as those asking changes to the \texttt{import} statements, which are ignored by previous techniques. Given also its ability to work with code written in nine different programming languages, this is currently considered the state-of-the-art technique.




\section{Study Design} \label{sec:design}

The \emph{goal} of this study is to assess the capabilities of state-of-the-art techniques for code review automation. The \emph{context} consists of: (i) three  techniques, \ie Tufano R. \etal \cite{tufano:icse2022} (\tufano), Li Z. \etal \cite{li:esecfse2022} (\li), and Hong \etal \cite{hong:fse2022} (\hong); (ii) two code review tasks for which the subject techniques provide automation, \ie \cnl and \cc; and (iii) instances featured in the test datasets on which the  techniques have been evaluated in the papers presenting them. We do not aim at comparing the performance of the three techniques to understand which one is the best. Rather, we look at them as a whole to understand the status of code review automation. 

We address the following research questions (RQs):\smallskip

\textbf{RQ$_1$: \emph{What are the characteristics of correct and wrong recommendations generated by techniques for code review automation?}} We cluster the predictions generated by the experimented techniques into two sets representing instances for which the approaches generated a correct or a wrong prediction. Then, we quantitatively compare these two sets. For the \cnl task we compare the ``complexity'' of the comments to automatically generate (\ie those present in the ground truth). Similarly, for the \cc task, we compare the ``complexity'' of the code changes to implement. Such an analysis will shed some light on the extent to which the state-of-the-art techniques overfit towards the low-hanging fruits of the datasets. 

On top of this, we qualitatively analyze \manually predictions generated by the three approaches (equally distributed between correct and wrong predictions) to characterize the type of code change they were able to request (\cnl task) or to automatically implement (\cc task). The objective is to understand the scenarios in which these techniques are successful \emph{vs} those in which they tend to fail. 
For example, if the outcome reveals that for the \cc task the techniques are mostly successful in implementing formatting changes, but tend to fail when dealing with more challenging code changes (\eg fixing a bug), this would question their usefulness. 

\textbf{RQ$_2$: \emph{To what extent are the datasets used to train and test techniques for code review automation suitable for such a scope?}} Through qualitative analysis we unveil the presence of problematic instances in the datasets used in the subject studies, calling for better dataset-cleaning pipelines. 

\textbf{RQ$_3$: \emph{How do techniques for code review automation proposed in the literature compare to state-of-the-art large language models?}} We compare  the three subject techniques with ChatGPT \cite{chatgpt} as representative of LLMs. Such a comparison informs the need to further invest in automating code review through specialized models rather than relying on general-purpose LLMs.


\subsection{Study Context}
We present the study context in terms of experimented techniques and datasets/predictions we analyzed.

\subsubsection{Techniques for Code Review Automation}
\label{sub:techniques}


\rev{Our work focuses on techniques aimed at \emph{imitating} human reviewers, thus automating the \cnl and/or \cc task. Based on our analysis of the literature, five approaches target these tasks: Tufano R. \etal \cite{tufano:icse2021}, \tufano \cite{tufano:icse2022}, \hong \cite{hong:fse2022}, \li \cite{li:esecfse2022}, and Li L. \etal \cite{li_lingwei:esecfse2022}. Tufano R. \etal \cite{tufano:icse2021} has been excluded since in their followup work \cite{tufano:icse2022} the authors showed the superiority of the newly presented technique (\tufano) as compared to their first attempt \cite{tufano:icse2021} in automating code review. Li L. \etal \cite{li_lingwei:esecfse2022}, instead, has been excluded after inspecting the replication package shared by the authors: We rely on the authors' replication packages to download (and then inspect) the predictions generated by their model on the test set. The replication package  for Li L. \etal \cite{li_lingwei:esecfse2022} featured 987 predictions for the 1,055 instances in the test set, casting doubts on the mapping between test instances and predictions.} Also, 7 of these predictions had one or more ``partial duplicate'' in the training set, meaning that the training set featured the same code instance (\ie the same code for which the technique had to automatically generate ``reviewer's comments'') of an entry in the test set with a different target message. While this is not a problem in principle, this plays a role in our qualitative evaluation, where we analyze whether the comments generated by these techniques are semantically equivalent to the expected one (despite they might use a different wording). Having the same code instance in the training set allows the approach by Li L. \etal \cite{li_lingwei:esecfse2022} to ``reuse'' the reviewer's comment from the training set, thus generating something that, for sure, will be meaningful. For these reasons we discarded this technique from our study. This left us with the three techniques. 

For the \cnl task, we consider predictions generated by all three approaches with: \tufano \cite{tufano:icse2022} being representative of DL-based techniques working at method-level granularity and not considering the code diff as an input; \hong \cite{hong:fse2022} being representative of IR-based techniques also working at method-level granularity; and \li \cite{li:esecfse2022} being representative of DL-based techniques working at ``diff hunk'' granularity and considering the code diff as an input. For the \cc task we only consider \tufano and \li, since \hong does not provide support for such a task.

\subsubsection{Datasets and Predictions}
\label{sub:datasets}
From the replication packages of the three techniques \cite{tufano:icse2022,hong:fse2022,li:esecfse2022} we collected the test sets used for their evaluation and the corresponding predictions. The size of the test datasets is reported in \tabref{tab:related}. There are two main differences among the datasets. The first is in the representation of the code submitted for review that is provided as input to the technique ($C_s$). While for \tufano \cite{tufano:icse2022} and \hong \cite{hong:fse2022} $C_s$ is a Java method, for \li \cite{li:esecfse2022} is a diff hunk. The second concerns the fact that \tufano and \hong only work with Java code, while \li supports nine languages, including Java. In our study we only considered Java instances for consistency and to simplify the following described manual analysis.


\subsection{Data Collection and Analysis}
\label{sub:analysis}

\subsubsection{RQ$_1$: Correct \emph{vs} wrong recommendations} To answer RQ$_1$ the first step is to classify the predictions by the three approaches as correct or wrong. We considered a prediction as correct if it represents an exact match (EM) with the target (\ie the expected output). This means that for the \cnl task the model generated a comment identical to the one written by human reviewers, while for the \cc task the model implemented a code change required by the reviewer exactly as the human contributor did. For each pair of technique and automated task, such a process resulted in the identification of the buckets of correct and wrong predictions reported in \tabref{tab:inspected} (columns ``\# correct (\%)'' and ``\# wrong (\%)''). While the EM metric has been used in the evaluation of the three techniques, we acknowledge that it has strong limitations, since it provides quite a strict definition of correctness. For example, an automatically generated natural language comment requesting the same code changes of the target comment with different wording is considered wrong. While this undermines a purely quantitative assessment of performance, in our study we use EM only as a mean to identify \emph{candidate} correct and wrong predictions. The predictions will undergo a manual analysis which, for example, considers correct generated messages being semantically equivalent to those posted by the human reviewers. 

\begin{table*}
\caption{Inspected instances}
\centering
\scriptsize
\begin{tabular}{lcrrrrrr}
\toprule
\textbf{Reference} & \textbf{Task} & \textbf{\# correct (\%)} & \textbf{\# wrong (\%)} & \textbf{Inspected correct} & \textbf{Inspected wrong} & \textbf{Valid correct} & \textbf{Valid wrong}\\ 
\midrule
\multirow{2}{1.5cm}{{\bf \tufano \cite{tufano:icse2022}}} & \emph{code\&comment-to-code}     & 2'363 (14.08\%) & 14'417 (85.92\%)   & 178 & 272 & 199 & 167\\
                                                                                         & \emph{code-to-comment}                & 354 (2.11\%)  & 16'426 (97.89\%)       & 200 & 227 & 169 & 189\\ 
\cmidrule{1-8}
\textbf{\hong \cite{hong:fse2022}}                                   & \emph{code-to-comment}                & 479 (2.85\%)  & 16'301 (97.15\%)       & 234 & 254 & 169 & 176\\
\cmidrule{1-8}
\multirow{2}{2.4cm}{\textbf{\li \cite{li:esecfse2022}}}         & \emph{code\&comment-to-code}     & 599 (27.15\%) & 1'607 (72.85\%)       & 197 & 317 & 197 & 176\\
                                                                      & \emph{code-to-comment}                & 0  (0\%)           & 1'611 (100\%)          & -      & 412 & 50  & 179\\
\bottomrule
\end{tabular}
\label{tab:inspected}
\end{table*}

\textbf{Qualitative Analysis.} The goal of the manual analysis was to characterize the type of code change the experimented techniques were (or were not) able to request (\cnl task) or to automatically implement (\cc task). For each of the above-described buckets of correct and wrong predictions (as identified via the EM analysis), we targeted the manual inspection of 167 valid instances, corresponding to a statistically significant sample with \rev{at least} a confidence level of 99\% and confidence interval of $\pm$10\% \rev{ for each bucket. The target of 167 instances was defined by computing a sample size ($SS$) calculation formula~\cite{Rosner2011} on the bucket having the largest ``population'' (\ie wrong predictions generated by \tufano for the \cnl task, with 16'426 instances):}
\rev{
{
\[
SS=\frac{\frac{z^2 \times p(1-p)}{e^2}}{1+(\frac{z^2 \times p(1-p)}{e^2 \cdot N})}
\]
}
}
\rev{\noindent where $p$ is the predicted probability of the observation event to occur, set to 0.5 when not known a priori (as in our case), $N$ is the population size, $e$ is the estimated margin of error ($\pm10\%$), $z$ is the $z$-score for a given confidence level (in our case, 2.58 for the 99\% confidence). As it can be seen from the formula, the larger $N$, the larger the sample size. Thus, using the largest ``population'' to compute the number of instances to inspect is a conservative choice, ensuring even better confidence when working on smaller buckets.} 

We use the term ``valid'' instances to account for the following scenarios. First, when inspecting a prediction falling in one of the \emph{wrong} buckets it is possible that we realize that the prediction is actually correct (\eg the comment generated/retrieved by the technique uses different wording as compared to the target, but it is semantically equivalent). In this case, while we inspected a \emph{wrong} instance, it will actually fall into the corresponding \emph{correct} bucket. Second, we discarded several problematic instances we found in the test sets of the experimented techniques. 
\rev{For example, we found instances in the \cc task for which, given the input code as context, it was impossible even for a human to understand the associated reviewer's comment (\ie what the reviewer was asking to change in the input code). Indeed, the reviewer's comment referred to a wider context (\eg parts of the code base not provided as input to the model), making the prediction impossible for the automated technique. These are problematic instances in the test set rather than failure cases of the technique, and we document them in RQ$_2$. 
In summary, an instance was considered ``valid'' if, given the input information (\ie $C_s$ for the \cnl task, and $\langle C_s, R_{nl}\rangle$ for the \cc task), it was possible for a human to understand the rationale for the related output: For the \cnl task, this means that the human evaluator was able to understand what the $R_{nl}$ comment to generate refers to (\ie what the problem in the submitted code $C_s$ spotted by the reviewer is); for the \cc task, the evaluator considers an instance valid if the changes resulting in the revised code $C_r$  to generate actually address the reviewer comment $R_{nl}$ posted for the submitted code $C_s$. The instances to inspect were randomly selected from each bucket until the target number of valid instances was reached.}

The columns ``Inspected correct'' and ``Inspected wrong'' in \tabref{tab:inspected} report, for each bucket, the number of instances we ended up manually inspecting to reach our target of 167 valid instances per bucket. Overall, we inspected \manually instances. Each instance has been independently inspected by two authors (from now on, evaluators) who were tasked with classifying the type of change to request (\cnl) or to implement (\cc). Five authors were involved in the manual analysis. \rev{On average, the authors have 13.4 years of programming experience (min=6) and 9.2 years of experience with Java (min=5), the language used in the inspected datasets. One of them holds a PhD in software engineering, and three more are currently pursuing a PhD in software engineering. One is a software engineer.}

The whole process was supported by a web app we developed that implemented the required logic and provided a handy interface to visualize the instance to label. For each instance, the evaluator was presented with: (i) the input provided to the approach; (ii) the generated prediction; and (iii) the expected output. As a result of the inspection, the evaluator could classify the instance or discard it as non-valid, providing an explanation as to why it was discarded. 

The classification required to assign the instance one or more labels describing the change (\eg \emph{refactoring $\rightarrow$ extract method}). Each evaluator was free to define their own label(s) \rev{(\ie open coding procedure)}, as they felt it was needed to properly describe the change: For this specific task, it was not possible to define upfront all possible labels, \rev{making card sorting \cite{coxon1999sorting} not suitable for our study.} Indeed, while there are taxonomies of issues identified during the code review process \cite{mantyla:tse2009,belleretal:msr2014,Pascarella:CSCW2018} their abstraction level is not suitable for our goal. For example, the taxonomy by M\"antyl\"a \etal \cite{mantyla:tse2009} includes a category named \emph{evolvability defects $\rightarrow$ structure} which is too coarse grained to investigate the automation capabilities of the subject techniques. To provide a concrete example, the taxonomy depicting the types of changes that the three techniques were (or were not) able to automatically request in the \cnl task (\figref{fig:taxonomy_cnl}) we have an entire tree dedicated to the recommendation of refactoring operations (which would fall under the \emph{evolvability defects $\rightarrow$ structure} taxonomy from \cite{mantyla:tse2009}). Our taxonomy includes concrete refactoring operations (\eg \emph{Extract method}, \emph{Change variable/constant type}), some of which are successfully recommended by the experimented techniques, while others consistently represent failing scenarios. The fine-grained nature of our taxonomy allows to observe these differences. 

The agreement among the authors was to define each label in the form \emph{parent $\rightarrow$ child}, where \emph{parent} was a coarse-grained description of the change while \emph{child} was a more specific, fine-grained description. New labels defined by an evaluator were made available in the web application to the other evaluators. While this goes against the notion of open coding, this allows to reduce the chance of multiple evaluators defining similar labels to describe the same type of change while not substantially biasing the process. The evaluator was also in charge of flagging instances in the wrong buckets as ``actually correct'' in case they felt that the prediction, while different from the target, was correct. 

The manual evaluation was performed in three rounds. A first round asked each evaluator to inspect 30 instances. This round resulted in a set of labels that has been inspected by the authors with the goal of merging similar labels and come up with a common strategy to categorize the instances in the next rounds. Then, a second round was performed in which the authors targeted the labeling of 30\% of the overall instances assigned to them. Again, such a round was followed by a further inspection of the defined labels, with grouping of similar labels and further discussion on strategies to improve agreement. \rev{The rationale for the number of instances to inspect in each of the three rounds was the following. We wanted to label very few instances in the first round (30) since we expected several inconsistencies in the way in which the authors were going to perform the labeling and, thus, we targeted a short dry run to test the adopted web application, the clarity of the overall process and, at least in part, the type of labels assigned by the authors (\eg their granularity). Then, we decided to follow with a larger second batch (30\%) which allowed to spot more corner cases worth to be discussed among the authors (\eg instances for which an author was unsure about the type of label to assign). Finally, since we felt that the labeling process was well-defined and clarified among the authors, we decided to move on labeling the whole dataset.} Once all \manually instances have been inspected by two evaluators, we solved conflicts that arose in \conflicts cases\footnote{Given the open nature of the coding, it was not possible to compute a meaningful inter-rater agreement (\eg Cohen’s kappa).}. Such a number may look high, since it represents 53\% of the inspected instances. However, the high rate of conflicts is explained by three design decisions. First, we considered all conflicts, also the ones resulting in the first and second round in which the set of possible labels was not stable at all. Second, the labels in our study emerged from the data and were not pre-defined. To get an idea of the complexity of this task the whole process resulted in a total of \labels different labels. Third, we were conservative in our definition of conflict, which occurred if: (i) two evaluators assigned a different set of labels to the instance, even if the two sets partially overlapped; (ii) two evaluators assigned the exact same set of labels to a \emph{wrong} instance with only one of the two reporting the instance as ``actually correct''; (iii) only one of the two evaluators labeled the instance, while the other one discarded it. Each conflict has been inspected by two additional authors, who discussed and solved it.  

Finally, we used the assigned labels to build hierarchical taxonomies showing the types of changes in which the three techniques tend to succeed and fail for the two automated tasks. Such a process required additional inspections of the considered instances. Indeed, once all categories in the taxonomies have been defined, two of the authors rechecked that all instances were assigned to the most proper category. Indeed, it is possible that a category $C$ added during the very last labeling round would be more suitable for instances inspected at the very beginning of the manual process, when $C$ was not available (since no one came up with that label while inspecting the instance). This resulted in the re-assignment of 16 instances ($\sim$1\%).

The final number of valid instances (\ie non-discarded) within each bucket is reported in the columns ``Valid correct'' and ``Valid wrong'' in \tabref{tab:inspected}. A few clarifications are needed for values which are different from 167, which was our original target. First, we did not have correct (EM) predictions generated by \li \cite{li:esecfse2022} for the \cnl task. Thus, we applied the following procedure to collect instances for the corresponding bucket. We selected among the \emph{wrong} predictions generated by \li, the top-100 in terms of BLEU-4 (Bilingual Evaluation Understudy) score \cite{papineni:acl2002}. BLEU measures how similar the candidate (predicted) and reference (oracle) comments are in terms of overlapping $4$-grams. \rev{A value of 1.0 indicates that the candidate and the predicted comment are identical. The selected top-100 predictions have a BLEU-4 ranging between 0.28 and 0.72.}

Our assumption is that \emph{wrong} predictions having a high BLEU score are likely to be correct, since they closely resemble the target comment. During the manual analysis, we discarded the instances that despite the high BLEU score, were actually wrong, since they did not belong to the ``correct bucket''. This process led us to 50 valid instances in this bucket, which is the only one being underrepresented (see \tabref{tab:inspected}). Second, as it can be seen from \tabref{tab:inspected}, in several buckets we collected more than the targeted 167 valid instances. This is due to the conflict resolution phase in which some instances discarded by one of the two evaluators were considered valid and re-introduced.

The output of this analysis are two taxonomies reporting, for each of the two tasks, the types of code changes on which the experimented techniques tend to succeed or to fail. 

\textbf{Quantitative Analysis.} We contrast the complexity of the test set instances resulting in correct and wrong predictions of the experimented techniques. For the \cnl task, we used as proxy for complexity the number of words featured in the comment to generate, under the assumption that longer comments are likely more complex. 

For the \cc task, we measure the number of AST-level changes required to convert $C_s$ (\ie the code submitted for review) into $C_r$ (\ie the revised code addressing the reviewer's comment). We expect a higher number of changes to indicate a higher complexity of the comment to implement. The AST-level changes have been extracted using Gumtree Spoon AST Diff \cite{Falleri:ase2014}. 

\rev{For the \cnl task, we used as proxy for complexity (i) the number of words featured in the comment to generate, under the assumption that longer comments are likely more complex, and (ii) the number of AST-level changes required to address the reviewer's comment (as done for the \cc task). The latter was only computed for the predictions generated by \tufano and by \hong, since for \li we did not manage to retrieve from the dataset the code implementing the required change, but only the submitted code with the posted reviewer's comment.}

For both tasks, we report boxplots of the distribution of the complexity proxies  for correct and wrong predictions both overall and by approach. We also statistically compare the two distributions assuming a significance level of 95\% and using the Wilcoxon test \cite{wilcoxon:ibs1992}. The Cliff's Delta ($d$) is used as effect size \cite{grissom:lawrence2005}, and it is considered: negligible for $|d| < 0.10$, small for $0.10 \le |d| < 0.33$, medium for $0.33 \le |d| < 0.474$, and large for $|d| \ge 0.474$ \cite{grissom:lawrence2005}. We adjust $p$-values using Holm's correction procedure \cite{holm:sjs1979}. We  compare the complexity proxies only on the predictions we manually validated. The reason is that, as explained, relying on EM to identify correct predictions could lead to false negatives, thus invalidating our quantitative analysis.  

\subsubsection{RQ$_2$: Datasets quality} The datasets used to train and test the experimented techniques have been automatically mined from GitHub. The authors applied a number of heuristics to filter-out problematic instances. For example, in the \cnl task efforts have been made to remove review comments posted by bots. Similarly, in the \cc task in which $\langle C_s,  R_{nl}\rangle~\rightarrow C_r$ triplets are involved, checks are performed to make sure that $C_r$ (the code which should implement the reviewer's comment $R_{nl}$) is different from $C_s$. Indeed, $C_s = C_r$ $\implies$ $R_{nl}$ not addressed.

Despite the effort in cleaning the datasets, in our manual analysis we found $\sim$25\% of the inspected instances representing noise in the datasets, posing questions on the reliability of the evaluations reported in the literature. As previously explained, when discarding an instance during the manual analysis the evaluators had to report the reason why said instance was discarded. After such a process, the five authors looked at the provided motivations and distilled them into four main categories representing errors introduced during the automated mining of the data from GitHub. We answer RQ$_2$ by presenting statistics summarizing such an analysis. 

\subsubsection{RQ$_3$: Comparison with LLMs}
\label{sub:rq3design} 
We assess the performance of ChatGPT \cite{chatgpt} on the two tasks focus of our study. 
We limited the number of samples to 250 for ChatGPT, due to the high cost of running such an evaluation. Indeed, there were two manual steps to perform. First, we needed to interact with the ChatGPT GUI to manually prompt each instance on which we wanted to run ChatGPT. Based on some tests we performed, we ended up selecting the following two prompts for our tasks:

\textbf{\cnl}:~ \texttt{Write a code review of the following code ``\{inputCode\}''}.\smallskip

\textbf{\cc}:~ \texttt{Revise this code ``\{inputCode\}'' given this comment ``\{inputComment\}''}.

In the prompts \texttt{\{inputCode\}} and \texttt{\{inputComment\}} represent the $C_s$ and $R_{nl}$, respectively, in the test datasets used for the two tasks. The replies provided by ChatGPT when using these prompts made it clear that it properly interpreted the task to perform. 

Second, once ChatGPT generates an answer, we cannot rely on EM to check if it is correct, since ChatGPT has not been trained to generate answers in the same format used in the test sets. For this reason we had to manually inspect the generated answers to assess their correctness. Each generated answer was independently inspected by two authors, who classified it as correct or wrong. For the \cnl task we considered the code review generated by ChatGPT as correct if it contains the target comment. For \cc, we verified whether ChatGPT properly addressed the reviewer's comment, even with the coding solution being different to the target one. Conflicts (\ie the two authors disagreed on the correctness of ChatGPT's answer), which arose in 18\% of cases, were solved by a third author.

For the \cnl task we randomly selected 50 instances from the test set of each of the experimented techniques (150 instances in total). The 50 instances included 25 correct (\ie the corresponding technique generated a correct solution) and 25 wrong predictions. The same approach has been used for the \cc task which, however, is only automated by two of the three subject techniques, thus resulting in 100 randomly selected instances.

We answer RQ$_3$ by reporting the percentage of cases in which ChatGPT was successful in both tasks. We also analyze the overlap between the state-of-the-art techniques and ChatGPT by reporting the percentage of cases in which (i) both succeed; (ii) at least one of the two succeeds; and (iii) none of the two succeeds.
\section{Results Discussion} \label{sec:results}

\begin{figure*}
	\centering
	\includegraphics[width=\linewidth]{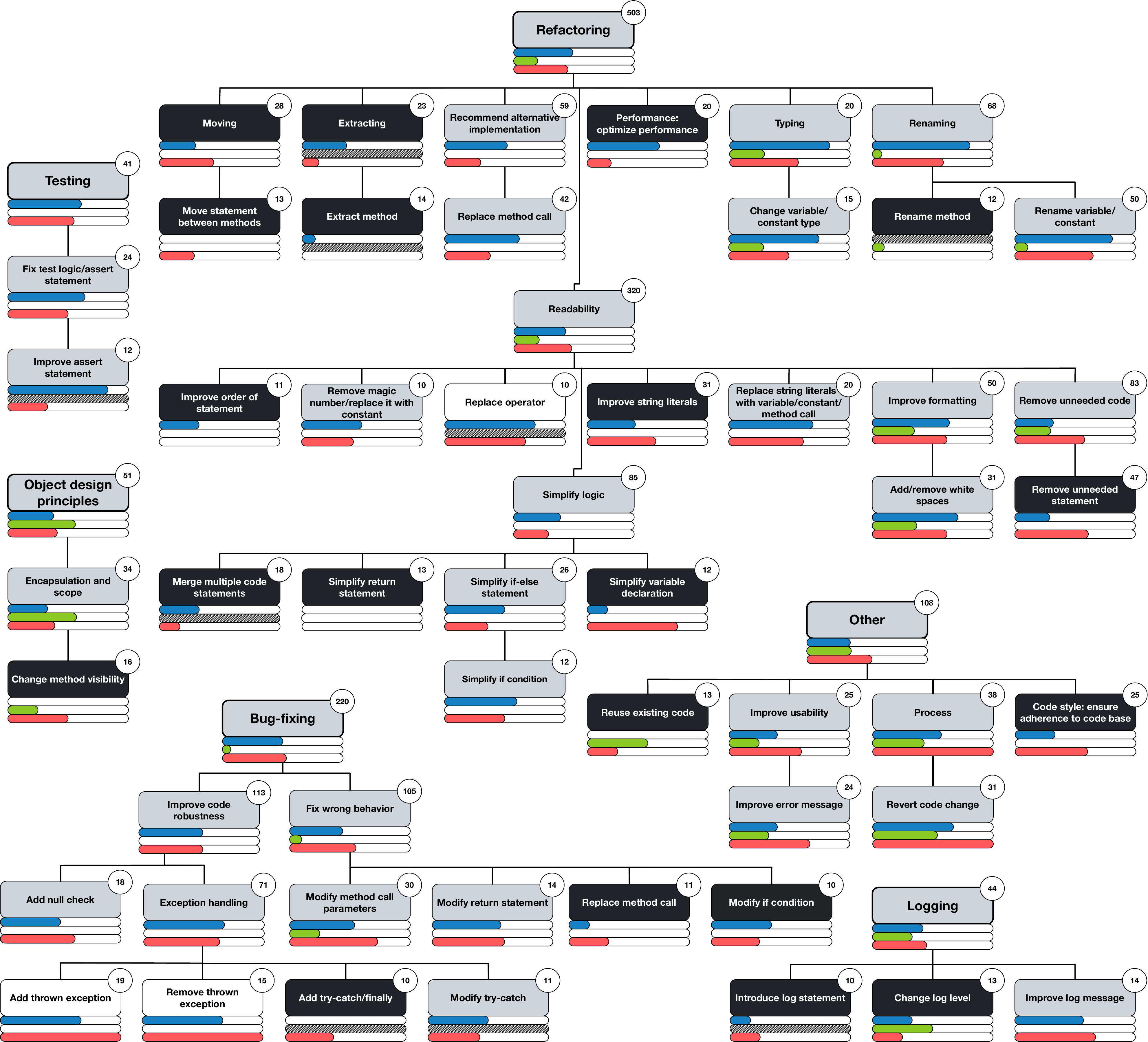}
	\caption{Taxonomy of types of changes for the \cnl task. The color assigned to each label reflects the ability of the techniques to automate the code review task in the context of such a change type (\fcolorbox{black}{white}{white} best, \colorbox{black}{\textcolor{white}{black}} worst). We report the percentage of successful predictions by each approach for each change type as bars below the corresponding category: \tufano (blue bar), \li (green), and \hong (red). 
	}
	\label{fig:taxonomy_cnl}
\end{figure*}

\begin{figure*}
	\centering
	\includegraphics[width=\linewidth]{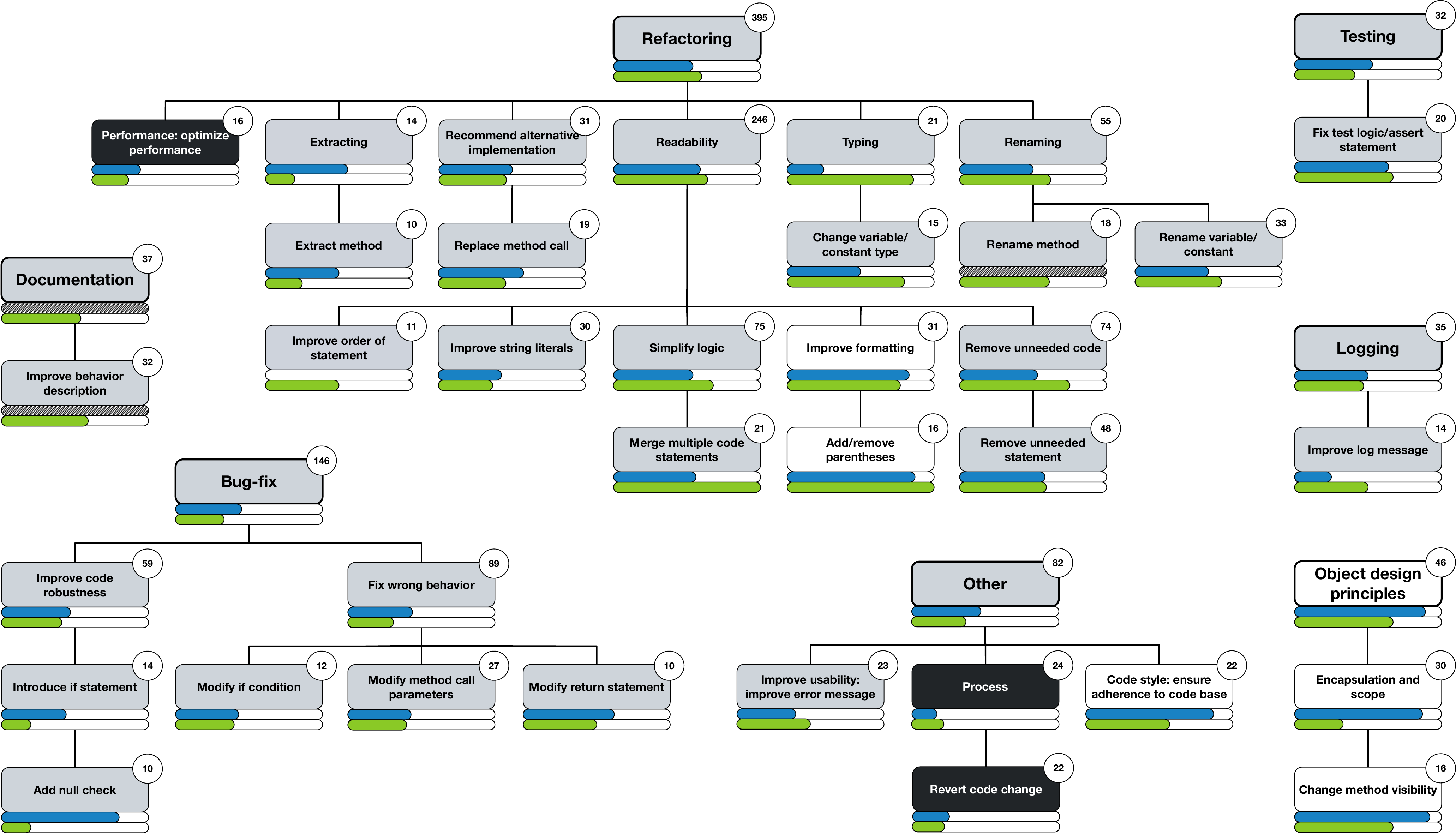}
	\caption{Taxonomy of types of changes for the \cc task. The color assigned to each label reflects the ability of the techniques to automate the code review task in the context of such a change type (\fcolorbox{black}{white}{white} best, \colorbox{black}{\textcolor{white}{black}} worst). We report the percentage of successful predictions by each approach for each change type as bars below the corresponding category: \tufano (blue bar), \li (green). 
	}
	\label{fig:taxonomy_cc}
\end{figure*}

We discuss the achieved results by RQ. We use the \faLightbulbO~icon to mark lessons learned and directions for future work.
\subsection{RQ$_1$: Correct \emph{vs} wrong recommendations}

\figref{fig:taxonomy_cnl} and \figref{fig:taxonomy_cc} report the taxonomies of ``types of code changes'' involved in the predictions generated by the subject  techniques. \figref{fig:taxonomy_cnl} refers to the \cnl task, depicting types of changes that the techniques were supposed to ask for in generated comments, as a human reviewer would do. \figref{fig:taxonomy_cc} refers to the \cc task, reporting types of changes that the techniques were required to automatically implement.

The taxonomies include several trees, each one representing a generic set of code changes specified in the root category (\eg \emph{refactoring}, \emph{bug-fix}). The number on the top-right corner of each label reports the number of instances we manually assigned to that change type (\eg 503 refer to \emph{refactoring} in \figref{fig:taxonomy_cnl}). 

Three clarifications are needed on this point. First, one instance we analyzed (\ie a prediction generated by an approach for a test entry) could have been assigned to multiple categories since requiring multiple types of changes. Second, for readability reasons, we decided to not report in the picture all categories of code changes that have been assigned to less than ten instances. Indeed, it is difficult to draw any conclusion with so few data points. The full data is available in our replication package \cite{replication}. 

Third, being a hierarchical taxonomy, one may expect the number of elements in a parent category to match the sum of the number of elements in its child categories.
\rev{However, this is not the case due to two reasons. First, sometimes the parent category has been used as a label when we did not manage to clearly identify the required code change, but only its overall goal (\eg using \emph{bug fix $\rightarrow$ fix wrong behavior} instead of its child \emph{modify if condition}). Second, as previously explained, we do not depict in the taxonomies categories featuring less than 10 instances. However, we still count their instances in the parent category (\eg \emph{refactoring $\rightarrow$ renaming} has a \emph{rename class} child which has not been depicted but contributes with 6 instances).}


In this scenario, the parent category has been used as a label when we did not manage to clearly identify the required code change, but only its overall goal (\ie fixing wrong behavior). This is another reason why parent categories can have a higher counting than the sum of their child categories.

The color assigned to each label reflects the ability of the techniques to automate the code review task in the context of such a change type. Since we manually analyzed $\sim$50\% of correct and $\sim$50\% of wrong predictions generated by each approach, a success rate around 50\% for a change type indicates that the techniques do not tend to perform particularly well or bad for that change type. 

Indeed, the goal of this analysis is to see if the correct (wrong) predictions are polarized by specific categories. For these reasons, we defined the color schema as follows:

A \colorbox{gray!35}{gray} label indicates a change type for which the automation level aligns with what expected based on our sample of correct and wrong predictions. Looking at \tabref{tab:inspected} it can be seen that, for the \cnl task, we inspected a total of 388 correct and 544 wrong instances. Such an imbalance is due, as previously explained, to the \li technique which generated 0 EMs and for which we decided to manually inspect 100 wrong predictions looking for actually correct ones (50 identified). This means that we should expect an average performance per change type close to (388*100)/(544+388)=42\%. For this reason, \figref{fig:taxonomy_cnl} features a grey category if the techniques succeeded for such a change type in 32\% to 52\% of cases (\ie 42\% $\pm$ 10\%). With a similar computation, \figref{fig:taxonomy_cc} features a grey category if the techniques succeed in 43\% to 63\% of cases (since an average of 53\% correct predictions has been analyzed per approach). Note that the ``$\pm$ 10\% choice'' has been done to simplify the results discussion and visualization. We acknowledge that other choices are possible (\eg $\pm$ 20\%); raw data with exact percentages are available in our replication package \cite{replication}.

A \colorbox{black}{\textcolor{white}{black}} label indicates a change type for which the automation tends to fail, thus in which the techniques are struggling. This means a success rate lower than 32\% for the \cnl task, and lower than 43\% for the \cc task.

A \fcolorbox{black}{white}{white} label indicates a change type for which the automation succeeds more than expected, namely in at least 53\% of cases for \cnl and 64\% of cases for \cc.

While this 3-level color schema represents the capabilities of the experimented techniques as a whole, \figref{fig:taxonomy_cnl} and \figref{fig:taxonomy_cc} also report the percentage of successful predictions generated by each approach for each change type as bars below the corresponding category. In \figref{fig:taxonomy_cnl} the three bars are ordered from top to bottom as: \tufano (blue bar), \li (green), and \hong (red). In \figref{fig:taxonomy_cc} the bars are only two, corresponding to  \tufano (blue) and \li (green). An empty bar indicates that the approach always failed for instances of that type. Instead, the filling of the bar with a zig-zag pattern indicates that the manually inspected test set entries on which the corresponding technique has been experimented did not contain any instance of that type. 


On top of the two taxonomies, \figref{fig:boxplot} depicts the boxplots showing the computed ``complexity'' of the test instances on which the experimented techniques succeed (blue) or fail (orange). We discuss our qualitative and quantitative results by automated task.

\subsubsection{Code-to-comment} Before looking at characteristics of correct and wrong predictions, \tabref{tab:nonEM} reports, for each approach and for each task, the percentage of non-EM predictions that we classified as actually correct. As it can be seen, 21.83\% of non-EM predictions generated by \li are actually correct for the \cnl task. The percentage is smaller for the other two techniques for which we did not focus the inspection on predictions having a high BLEU, but still non-negligible ($\sim$2.5\%). \rev{For example, in the case of \hong the EM predictions are 2.85\% of the test set instances, while we found an additional 2.22\% of non-EM predictions which are actually correct, almost doubling the approach's correctness. A manual analysis of (a sample of) non-EM predictions is needed to better assess the capabilities of an automated technique.} 
An example of non-EM generated by \li and being actually correct belongs to the \emph{other $\rightarrow$ reuse existing code} category: The target comment was ``Use \texttt{IOUtils} instead'', while \li generated the equivalent ``Can we use Guava's \texttt{IOUtils} here?''. This is a first important outcome of our study: \faLightbulbO~ Using EM to assess the automation of the \cnl task might be unfair.

\begin{table}
\caption{Non-EM classified as correct during manual analysis}
\centering
\scriptsize
\begin{tabular}{lcr}
\toprule
\textbf{Reference} & \textbf{Task} & \textbf{\% correct non-EM}\\ 
\midrule
\multirow{2}{1.5cm}{\textbf{\tufano}} &\cc       & 15.66\% \\  
                                                         & \cnl     &   2.58\% \\     
\cmidrule{1-3}
\textbf{\hong}                                    &\cnl      &   2.22\% \\      
\cmidrule{1-3}
\multirow{2}{1.5cm}{\textbf{\li}}         & \cc      & 17.37\% \\   
                                                         & \cnl     &  21.83\% \\  
\bottomrule
\end{tabular}
\label{tab:nonEM}
\end{table}

\rev{Not surprisingly the white categories (\ie the techniques tend to succeed) are characterized by simple requests to include in the generated message, and in particular the \emph{removal/\-addition of a thrown exception}, and the \emph{replacement of an operator}.}
The excellent performance achieved in these change categories are usually driven by the success of the \hong IR-based technique. The latter has 100\% accuracy in recommending the addition/removal of exceptions, thanks to the retrieval from the training set of past reviewers' comments requiring such a change for similar methods. \faLightbulbO~ Looking at the taxonomy, it is clear that for code change types which are quite general, simple, and thus likely to be requested over and over again in different code review instances (\eg the addition/removal of exceptions, asking to \emph{revert a code change}), an IR-based approach can be a trump card. Differently, comments requiring the description of more complex changes are challenging to retrieve or synthesize. The \emph{refactoring} tree provides interesting examples. 

\begin{figure}
	\centering
	\includegraphics[width=0.99\linewidth]{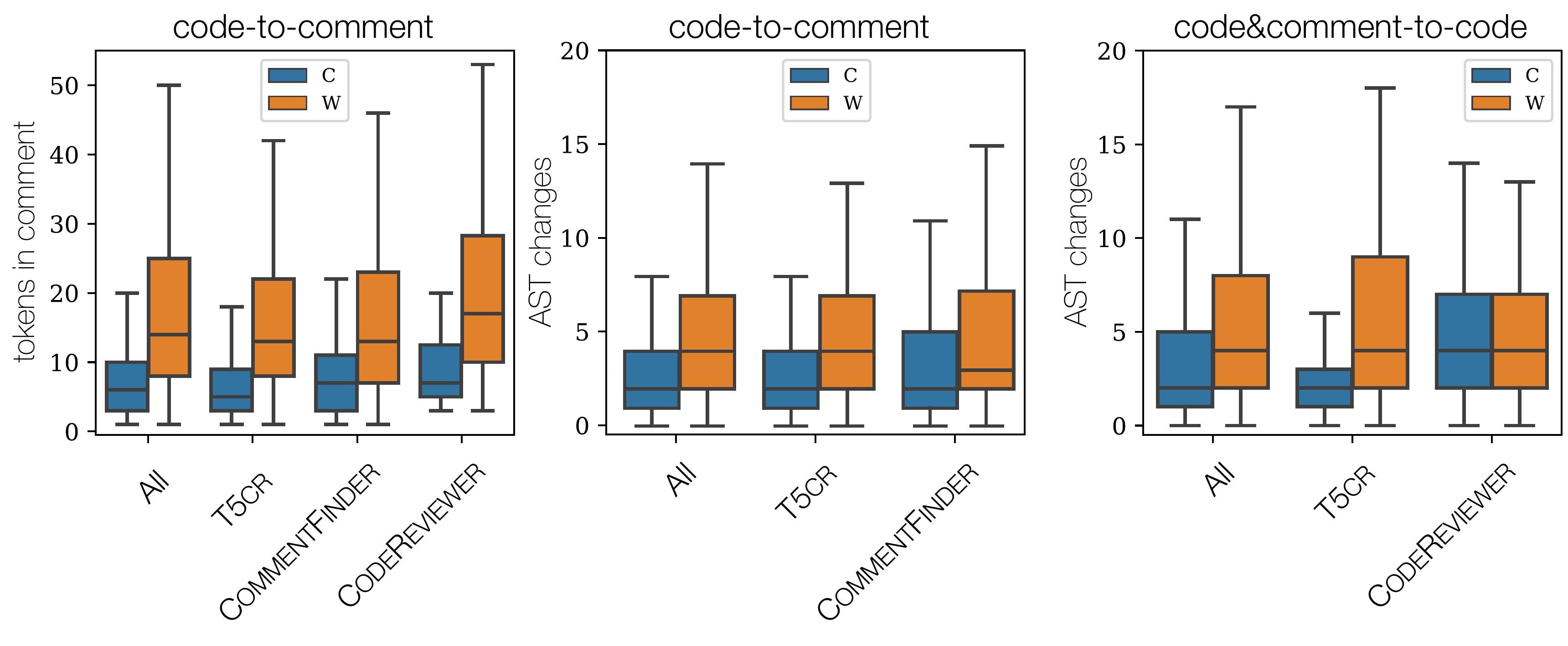}
	\caption{Task complexity for correct and wrong predictions}
	\label{fig:boxplot}
\end{figure}

Simple refactorings such as \emph{changing variable/constant type} or \emph{renaming variable/constant} are overall well-supported (\eg~\hong: ``\texttt{qry} $->$ \texttt{query}''). When it comes to refactorings involving complex code changes, possibly impacting multiple code components, the techniques tend to fail. This is the case for refactorings \emph{extracting} or \emph{moving} code elements. This is likely due to the limited contextual information provided to code review automation techniques. Among the experimented techniques, \tufano and \hong work at method-level granularity, meaning that the method provided as input represents everything the model knows about the system. Similarly, the ``view'' of \li is limited to a diff hunk. Such an issue does also affect the performance of the techniques for apparently simple changes to require, such as those asking to \emph{change the method visibility}. Indeed, without additional knowledge of the system it is difficult to judge what the correct visibility of a method should be. \faLightbulbO~ Pushing the boundaries of code review automation for these types of changes requires enriching the contextual information provided to the techniques, similarly to what observed for other software engineering tasks \cite{tian:icsme2022}.

A negative exception in the refactoring tree is the \emph{renaming of methods} which one could expect to be on a similar level of difficulty as compared to the \emph{renaming of variable/constants} which is, instead, quite successful. We inspected these instances and we noticed that, while renaming variables/constants may just require a term expansion (as in the \texttt{qry} example previously reported), methods' names are more expressive and challenging and, while the techniques are sometimes able to capture the need for a renaming, they fail in recommending meaningful alternatives. 

The techniques also have a hard time automating logging activities, especially when it comes to suggest the \emph{introduction of a log statement} or the need to \emph{change the log level} (\eg from \texttt{error} to \texttt{warning}). Interestingly, for both method renaming and recommendation of log-related changes, specialized DL-based techniques have been proposed in the literature (see \eg Alon \etal \cite{alon2019code2vec} for renaming, and Mastropaolo \etal \cite{mastropaolo:icse2022} for logging). Based on the reported empirical evaluations, those techniques proven to be quite effective in these tasks. For example, Mastropaolo \etal \cite{mastropaolo:icse2022} presented LANCE, an approach able to correctly recommend fixes to the level of log statements in 66\% of cases. \faLightbulbO~ While the code review automation techniques proposed in the literature target the automation of generic code changes, the adoption of specialized techniques might be more effective for specific change types. \rev{However, this might not be straightforward to do in the context of the \cnl task. Indeed, assuming the will to specialize a model for ``commenting'' on a specific type of issue, the first needed ingredient is a training dataset, which might not be easy to collect. One may think to cluster reviewers' comments  via lexical analysis and train a specialized model on each of those clusters. Nevertheless, a trade-off between cohesiveness of the clusters and availability of training data soon becomes evident: very cohesive clusters will result in highly specialized models which, however, are likely to benefit from very little training data (\eg only a few instances in which a reviewer's comment is suggesting to introduce a log statement). Larger clusters featuring more training data are instead unlikely to specialize the model for specific types of recommendation, thus again pushing it towards a generic recommender. A more promising approach might be to manually define ``commenting patterns'' for a specific type of change (\ie a standard sentence expressing the need for improving a certain aspect of the code, such as introducing a log statement). In this case, the training dataset could be built by parsing code changes performed during the change history of a project (\eg commits introducing a new log statement), independently from the availability of code review information for these changes. This implies the possibility to reliably identifying code changes in which the target issue  has been fixed. For some of the ``black categories'' in our taxonomy this can be easily achieved (\eg lack of log statement, need for changing the method visibility, \etc). Others would require more advanced solutions, like the usage of tools to detect refactoring operations \cite{Tsantalis:tse2022}. Negative instances, \ie code components on which the target issue does not manifest (\eg no need to add log statements), might be needed as well. Once trained, specialized models can be triggered on the code change submitted for review, reporting the improvement recommendations (if any) to the developer.}

Not surprisingly, the experimented techniques do not shine in recommending types of code changes which are likely to be system-specific and, thus, difficult to learn/retrieve from other sources. This is the case for the \emph{performance optimizations} comments that the techniques were required to emulate (\eg {\scshape~Target}: ``We should use \texttt{keyService} here, intention is to cache key temporary so under heavy load we don't download keys all the time''). \faLightbulbO~ A possible strategy to overcome this limitation could be to fine-tune the techniques to a specific software project or, at least, to a set of projects falling in the same domain (\eg DB engines). For example, after a pre-training performed on generic code review data, a DL-based approach could be fine-tuned to specifically support code review in a project. A major obstacle  is the possible lack of fine-tuning training data, since a single project is unlikely to provide enough training instances. This may be partially overcome through data augmentation techniques \cite{Shiwen:jss2022}.

\tufano and \hong achieve good performance when it comes to asking for \emph{testing}-related changes, correctly generating comments aimed at both improving the test coverage/logic (\eg~\tufano: ``add a check here to verify that the \texttt{serial\-Data\-Received method} was not called'') and cleaning/refining them, \eg~\tufano suggested to replace an empty \texttt{String} passed as parameter in an assert statement with an \texttt{EMPTY\_VALUE} constant already used in other statements of the test. In general, the changes to recommend in the \emph{testing} category are very specific and tend to focus on a single code statement.\\ \faLightbulbO~Still, this shows the potential of these techniques as possible ``refinement tools'' for approaches supporting the automated generation of test cases \cite{PachecoE2007Poster,Fraser:fse2011}.  

From the quantitative perspective, the boxplots in \figref{fig:boxplot} for the \cnl task suggest, as expected, that the techniques tend to succeed in the generation of simple reviewers' comments, having a median of 6 words composing them. As a comparison, the failing cases are more than twice longer (in terms of median), with 14 words. Such a trend holds, with minor differences, for all approaches. Also, it is interesting to notice that, when considering all techniques together, the first quartile of the wrong predictions is very close to the third quartile of the correct predictions, indicating a strong difference between the two sets that is confirmed by the statistical analysis with a $p$-value$<$0.001 accompanied by a large effect size (test results in \cite{replication}). \rev{Similar observations can be drawn when using the AST changes required to implement the reviewer's comment as a complexity proxy.} \faLightbulbO~ Future work should focus on boosting performance in these challenging scenarios, since the approaches seems to work pretty well in the generation of simple comments ($<$20\% of wrong predictions have less than 6 words).

\subsubsection{Code \& comment-to-code} Also for this task, we start by observing the substantial percentage of non-EM predictions which are actually correct --- 15.66\% for \tufano and 17.37\% for \li. \faLightbulbO~ This reinforces the need for manual analysis when assessing the performance of techniques for code review automation.

The taxonomy depicted in \figref{fig:taxonomy_cc} is smaller as compared to the previous one, since a higher number of categories (91) count less than 10 instances (for the full taxonomy see \cite{replication}). It is interesting to see some major differences as compared to the previous taxonomy. Changes which were trivial to ask for in a comment to generate (\eg ``please revert this change'') are challenging to automatically implement, as required in the \cc task. Indeed, the reverting may require several code changes which are not necessarily easy to predict, especially if the full code diff is not part of the information available to the model. 

Interesting is the complementarity between the two techniques that support this task.~\tufano is very effective in changes related to \emph{object design principles}, \eg handling issues related to \emph{encapsulation and scope} of variables/methods, which usually require minor code changes. Also, \tufano works  well in implementing changes \emph{ensuring adherence to the code base} in terms of \emph{coding style}, \eg addressing comments like ``\texttt{String.isEmpty()} is used in other places'' pointing to an inappropriate \texttt{if} condition checking \texttt{coverageId.length() == 0}.~\li, instead, is the only approach supporting documentation changes, and can achieve excellent performance even for less-trivial code changes requiring \eg to \emph{merge multiple code statements} (in order to improve readability) or to migrate towards more appropriate types (\emph{refactoring} $\rightarrow$ \emph{typing}).  Considering that both techniques are built on top of a transformer-based architecture, such a complementarity can be partially explained by the different code representation they use, one looking at a single method at a time (\tufano) and one taking a diff hunk into account (\li), possibly with a partial view of specific code components (\eg only the changed lines of a method are visible in the diff hunk).\\ \faLightbulbO~ A hybrid representation including both the full representation of the involved code entities and the changed lines might help in getting the best of both worlds.

As expected, both approaches are effective in the automation of very simple changes related to \emph{improve the formatting} of code (\eg \emph{add/remove parentheses}, \emph{add/remove white spaces}). The automation of these changes can be easily performed by a code formatter (\eg \cite{prettier}), without the need for expensive DL models. \faLightbulbO~ Such instances should be removed from the test sets used in the evaluation of techniques automating the \cc task, to avoid inflating the percentage of EM predictions they generate.

The two techniques struggle to automatically implement complex \emph{bug-fixes} requiring major code changes (\eg \emph{fix wrong behavior}) and/or changes to the code logic (\eg \emph{modify if condition}). \faLightbulbO~ This result is inline with what observed for techniques specialized in automated bug-fixing \cite{tufano:tosem2019}, which also tend to be successful in a minority of cases (usually $<$30\%) confirming the need for more work in the area.


Finally, we want to comment on the performance of the two techniques on the 87 types of code changes which are not represented in Figure 2, since counting less than 10 instances each. Overall these 87 categories feature 132 of the predictions we inspected for the \cc task.

\rev{Out of those, 69 (52\%) were correct predictions, which matches the expected success rate and seems to suggest that the two state-of-the-art techniques do not really struggle in automatically implementing code changes which are likely to be less represented in the training set. However, by inspecting the predictions in these categories, we found out that a ``good'' level of performance (\ie $\geq$52\% correct predictions) is only obtained for 48\% of these poorly represented categories (\eg for 18 of them we observed 0\% of correct predictions). It is thus possible that, in some cases, the models learn from similar and related categories in a sort of transfer learning fashion. For example, training on instances of the well-represented category ``\emph{remove unneeded statement}'' might have played a role in achieving good performance on the five instances belonging to the code change type ``\emph{remove final modifier}''. However, given the low number of instances in each of these categories (at most 9), we cannot draw any conclusion based on these findings.}

The results of the quantitative analysis (right side of \figref{fig:boxplot}) show an interesting trend: while the correct predictions by \tufano require substantially simpler changes as compared to the wrong predictions (median of 2 AST changes \emph{vs} 4, $p$-value $<$ 0.001 with medium effect size), this is not the case for \li. Here the two sets are basically equivalent in terms of code change complexity (negligible effect size). Looking at the boxplots it is clear that \tufano tends to overfit to simpler changes, while \li, possibly thank to the diff hunk representation, is able to cope with more complex code changes as well, supporting its status as state-of-the-art approach.

\begin{table}
\caption{Categories of discarded instances}
\centering
\resizebox{\linewidth}{!}{
\begin{tabular}{p{2.2cm}rrrrrr} 
\toprule
\multirow{2}{*}{\textbf{Reason}} & \multirow{2}{*}{\textbf{\#}} & \multicolumn{2}{c}{\textbf{\tufano}} & \textbf{\hong} & \multicolumn{2}{c}{\textbf{\li}}  \\ 
\cmidrule(r){3-4}\cmidrule(r){5-5}\cmidrule(r){6-7}
                                                   &                                         & C\&NL2C & C2NL                           & C2NL              & C\&NL2C & C2NL\\
\midrule
Unclear comment                      & 281                                  & 32             & 55                               & 97                    & 57             & 40\\
No change asked                      & 182                                  & 24             & 13                               & 44                    & 30             & 71\\
Ignored comment                      & 74                                    & 25             &  0                                & 0                      & 49             & 0  \\
Wrong linking                            & 16                                    & 1               & 0                                 & 1                      & 4               & 10\\
Other                                         & 16                                    & 2               & 1                                 & 1                      & 1               & 11\\
\bottomrule
\end{tabular}
}
\label{tab:discarded}
\end{table}

\subsection{RQ$_2$: Datasets quality}
\tabref{tab:discarded} reports the number of instances that we discarded as being problematic together with the reason why they have been discarded. For the sake of space, we shortened the \cc task as C\&NL2C and the \cnl as C2NL. While \tabref{tab:discarded} shows also the results by test set of each technique, we focus the discussion on the overall trend (\#). The ``Other'' category contains 16 instances discarded for various but rare reasons, that we do not discuss but document in \cite{replication}. The remaining ``discarding reasons'' are sorted based on their frequency from top to bottom. 

For 281 cases, we assigned the \emph{unclear comment} label to discard the instance since it was impossible even for a human to understand what to implement (\cc) or what the target comment to generate was actually asking to the developer (\cnl). For the \cc task, this was due to the limited contextual information provided to the model (\ie the input). 
For the \cnl task, a recurring problem is again the lack of context but, this time, related to the conversation that happened between the contributor and the reviewer(s), which is not visible to the techniques.  For example, a comment saying ``ah, ok, that would be clearer'' is meaningless without knowing the previous exchanged messages. \faLightbulbO~As already observed in RQ$_1$, increasing the contextual information is a must to push the boundaries of code review automation.   

In 182 instances we inspected the reviewer's comment was not requesting any change. For the \cc task, this means that the approaches could not really address the comment by modifying the code (\eg ``Awesome work so far, Eli!''). For the \cnl task this means training and testing the technique for the generation of comments which are uninteresting given the automation goal. Indeed, these techniques aim at generating comments asking for code changes targeting the improvement of source code. Thus, comments like ``I am not sure what GitHub wants to tell me with this icon here :)'' should not be considered relevant for these approaches. The cleaning pipelines employed in the works presenting the experimented techniques fail in filtering out those meaningless instances.

In 74 cases (all related to the \cc task), while the reviewer's comment was asking for a change, the contributor was changing the code but not to address the comment. These instances penalize both the learning and the evaluation of the models. Indeed, even assuming that the model correctly implements the change required by the reviewer, during training the weights of the network will be revised to steer the prediction towards a different (wrong) target and, during evaluation, any quantitative metric is likely to point to a wrong prediction.

Finally, 16 instances result from errors while mining the dataset, since the code has been linked to a wrong code, \eg ``there's no need for \texttt{final} in interfaces'' for an input code not being an interface.

\faLightbulbO~Worth noticing is the overall number of discarded instances (574) out of the \manually manually analyzed (25\%). Since the test set is just a random selection of 10\% of data, we can assume a similar distribution in the training sets of the subject techniques. Thus, all discussed instances have the potential to hinder the training and bias the testing, since it is unreasonable to expect them to result in a correct prediction given the target. Supervised or unsupervised techniques aimed at removing these problematic instances are needed to make a further step ahead on code review automation thanks to higher-quality datasets.

\subsection{RQ$_3$: State-of-the-art \emph{vs} ChatGPT}

\begin{table}
\centering
\caption{Manual analysis of ChatGPT predictions}
\resizebox{\linewidth}{!}{
\begin{tabular}{lcccc}
\toprule
	\multirow{2}{*}{\textbf{Task}} &  \multicolumn{2}{c}{\faCheck\textbf{State-of-the-art}} &  \multicolumn{2}{c}{ \faRemove\textbf{State-of-the-art}}\\ 
	                                              &  \faCheck\textbf{ChatGPT} &  \faRemove{\textbf{ChatGPT}} 
	                                              &  \faCheck\textbf{ChatGPT} &  \faRemove{\textbf{ChatGPT}}\\
\midrule
        \cc  & 66\% & 34\% & 44\% & 56\% \\ 
        \cnl & 19\% & 81\% & 7\%   & 93\% \\ 
\bottomrule
\end{tabular}
}
\label{tab:chatgpt}
\end{table}

\tabref{tab:chatgpt} shows the results of the manual analysis assessing ChatGPT in automating code review: For each task (rows) we report the percentage of cases in which ChatGPT succeeds (\faCheck) or fail (\faRemove) for instances on which the state-of-the-art (SOTA) techniques were correct or wrong. Results are aggregated for  the three approaches, with raw data available in our replication package \cite{replication}.

ChatGPT performs slightly better than the SOTA in the \cc task, being able to address the reviewer's comment in 55/100=55\% of cases, as compared to the 50\% of the three techniques (we selected half instances on which they work, and half on which they fail --- see \secref{sub:rq3design}). Interesting is the complementarity between ChatGPT and the SOTA: ChatGPT succeeds in 44\% of cases in which the SOTA techniques fail. These are mostly instances in which the reviewer's comment provides little information about the change to implement. 
For example, ChatGPT addressed a reviewer's comment pointing to a ``\emph{wrong formula}'' in the code (``wrong formula'' is the full content of the reviewer's comment) by applying the following change: \texttt{forward * strikesLike.\-get(i) + shiftOutput;} $\rightarrow$ \texttt{forward * Math.\-exp\-(strikesLike.\-get(i)) + shiftOutput;}. This is a failing instance for the SOTA.

Concerning the \cnl task, ChatGPT succeeds in 19/150=13\% of cases, being less performant than the SOTA. Only 5 of those instances are failure cases for the SOTA. For this task the output of ChatGPT is not a single comment (as for the SOTA techniques) but a list of observations regarding the submitted code. In most of cases, we found these comments to be meaningful, but they often miss or are in contrast with the point raised by the human reviewer. 

This is the case for an instance in which the diff hunk reported a change from \texttt{trace} to \texttt{info} for the level of a logging statement. While the reviewer complained about this change (and the SOTA technique agreed with the human reviewer), ChatGPT was in favor of it, commenting: ``\emph{an info level logging statement would be more applicable}'').

\revs{Since there is evidence in the literature about the major role played by the provided prompt on the output produced by large language models when dealing with code-related tasks \cite{Mastropaolo:icse2023}, we further investigated whether it is possible to improve the performance of ChatGPT by exploiting different and more advanced prompts. We took the failure cases of ChatGPT (\ie  131 cases for the \cnl task and 45 cases for the \cc task) and we provided them again as input to ChatGPT, but with different promptings also exploiting information from our taxonomy. In particular, our taxonomies feature five root categories classifying the types of changes reviewers usually ask to implement in a code review process: \emph{Bug-Fixing}, \emph{Object-Design Principles}, \emph{Testing}, \emph{Refactoring}, \emph{Logging}. On top of that there is the \emph{Other} category grouping change types which cannot be attributed to any of the five above categories. We use these root categories to define two prompts (one for each of the two tasks), inspired by the chain-of-thought prompting methodology \cite{wei:nips2022} which has been successfully applied in natural language processing:}\vspace{0.2cm}

\noindent \revs{\textbf{\underline{\cc task}}}\\
\small
\revs{\texttt{Given this code ``\{inputCode\}'' and this comment ``\{inputComment\}'' and assuming you are an expert developer:}}

\begin{enumerate}
\item \revs{\texttt{Identify/Classify the type of code changes requested in the comment. Answer with one or more of the following:}}

\begin{itemize}
\item \revs{\texttt{Changes have been required to refactor the code to improve its quality;}}
\item \revs{\texttt{Changes have been required  since tests for this code must be written;}}
\item \revs{\texttt{Changes have been required  to better align this code to good object-oriented design principles;}}
\item \revs{\texttt{Changes have been required  to fix one or more bugs;}}
\item \revs{\texttt{Changes have been required  to improve the logging of its execution;}}
\item \revs{\texttt{Changes have been required  for other reasons not listed above.}}
\end{itemize}

\item \revs{\texttt{Implement the required code changes.}\\}

\end{enumerate}

\normalsize
\noindent \revs{\textbf{\underline{\cnl task}}\\}
\small
\revs{\texttt{Given this code ``\{inputCode\}'' and assuming you are an expert code reviewer:}}

\begin{enumerate}
\item \revs{\texttt{Decide whether the code needs to be revised or not. Answer True or False.}}

\item \revs{\texttt{If the response to the above point is True, then identify/classify the type of code change(s) required. Answer with one or more of the following:}}

\begin{itemize}
\item \revs{\texttt{Changes are needed to refactor the code to improve its quality;}}
\item \revs{\texttt{Changes are needed since tests for this code must be written;}}
\item \revs{\texttt{Changes are needed to better align this code to good object-oriented design principles;}}
\item \revs{\texttt{Changes are needed to fix one or more bugs;}}
\item \revs{\texttt{Changes are needed to improve the logging of its execution;}}
\item \revs{\texttt{Changes are needed for other reasons not listed above.}}
\end{itemize}

\item \revs{\texttt{Write a code review (\ie explain the changes to be performed, if any) based on your answers to the questions above.}\\}

\end{enumerate}
\normalsize

\revs{As it can be seen, both prompts guide the large language model towards the solution of the task without, however, giving it any extra information besides what would be available during the review process.}

\revs{As for the \textbf{\cc task}, we also experimented an additional prompt, which classifies the reviewer's comment into one of the categories in our taxonomies before asking the model to automatically implement the code changes required to address the comment:\\\\}
\small
\revs{\texttt{Assume you are an expert developer. Given this code ``\{inputCode\}'' implement the code changes requested by a code reviewer in this comment ``\{inputComment\}'', considering that the change requested is asking to [category from our taxonomy to which the comment has been manually assigned].}\\}
\normalsize

\revs{For example, assuming that the assigned category in our taxonomy \emph{logging}, the prompt will end with: \small \texttt{considering that the change requested is asking to improve the logging of its execution.}} \normalsize

\revs{In other words, we are simulating the scenario in which, besides providing a comment explaining the change to perform, the reviewer also provides a label classifying the type of code change required. Note that such a prompting cannot be experimented in the \cnl task, since in that case the reviewer comment is the output of the model rather than the input.}

\begin{table}
\centering
\scriptsize
\caption{\revs{Failure cases made successful via more advanced prompting}}

\begin{tabular}{llr}
\toprule
	\textbf{Task} &  \textbf{Prompt} &  \faCheck{\textbf{ChatGPT (\%)}} \\

\midrule
        \cc  & Chain-of-Thought & 23/45 (51\%) \\ 
        \cc & Label-Aware & 25/45 (55\%) \\\midrule
        \cnl & Chain-of-Thought & 13/131 (10\%) \\ 
\bottomrule
\end{tabular}

\label{tab:chatgpt2}
\end{table}

\revs{As previously done, two authors independently assessed the correctness of ChatGPT output, and conflicts have been solved by a third author. \tabref{tab:chatgpt2} reports the achieved results. For the \cc task, depending on the used prompt, we managed to make ChatGPT correctly predicting up to 25 of the previously identified 45 failure cases. It is quite interesting to see that just providing the model with an explicit label taken from our taxonomy and describing the type of change required in the reviewer's comment substantially helps the model in correctly implementing the required changes. As for the \cnl task, the chain-of-thought prompt helped the model only 13 of the 131 failure cases, confirming that generic large language models not specifically trained for this task struggles in dealing with it.}

\revs{For completeness, we also provide in our replication package \cite{replication} two files mapping the ChatGPT success and failure cases to categories in our taxonomy. We do not discuss the findings in terms of coverage of the different categories in our taxonomy since the number of instances we analyzed in RQ$_3$ is quite low, resulting in few data points for each category (\ie it is difficult to judge whether CHatGPT works well for specific code change types).}

\faLightbulbO~ChatGPT is a \revs{very} competitive baseline for the \cc task, \revs{with prompting playing a major role in boosting the model's performance}. However, it is important to consider the fact that ChatGPT may have seen the code addressing the reviewer's comment during training, questioning the extent to which such a comparison is fair. When it comes to generating reviewers' comments, SOTA techniques are superior, supporting the worth of further research in this direction.
\section{Threats to Validity} \label{sec:threats}

{\bf Construct validity.} In RQ$_1$ we classified the change type based on what was visible in the test set instance. For example, we did not see the conversation between the contributor and the reviewers, which could allow to better understand the reviewers' requests. However, this was done by design to identify ``unclear'' instances in RQ$_2$. 

Also, we considered the predictions of the three techniques on the test sets used in the original papers presenting them. This means that they have not been assessed on the same dataset. However, as previously explained, our goal was not to compare the capabilities of the three approaches, but rather to look at their strengths and weaknesses as representative of the state-of-the-art in code review automation.

{\bf Internal validity.} There are possible subjectiveness issues in the manual analyses. We mitigated this threat by always involving multiple authors inspecting the same instance. Still, as for any manual process, imprecisions are possible.


{\bf External validity.} In terms of studied techniques, we considered all those focusing on the automation of the two targeted tasks; we only excluded the approach by Li L. \etal \cite{li_lingwei:esecfse2022} for the reasons explained in \secref{sub:techniques}. All our RQs required manual analysis, resulting in limitations on the number of data points collected. 
\rev{Given the expense of the manual investigation, we targeted the inspection of a sample of instances ensuring at least a confidence level of 99\% and a confidence interval of $\pm$10\% on each of the analyzed bucket of predictions. More conservative choices (\eg 95\%$\pm$5\% confidence) would have required a much higher number of inspected instances (more than twice).}
Replications are needed to corroborate/revise our findings.
\section{Conclusion and Future Work} \label{sec:conclusion}

We assessed the capabilities of three state-of-the-art techniques for code review automation \cite{tufano:icse2022,hong:fse2022,li:esecfse2022}. Differently from the mostly quantitative evaluations available in the literature our study has a strong qualitative focus. Our study disclosed the scenarios in which state-of-the-art approaches tend to succeed and fail (RQ$_1$) and identified major issues in the quality of the datasets used for their training and evaluation (RQ$_2$). Finally, we showed that ChatGPT \cite{chatgpt}, as representative of Large Language Models, is a competitive technique for code review automation, but still struggles in several scenarios, justifying the need for more research on models specialized for such automation (RQ$_3$). 

\rev{Future work will focus on assessing (i) the usefulness of code review automation from the practitioners' perspective, and (ii) the code review automation opportunities offered by other state-of-the-practice techniques (\eg Copilot \cite{copilot}).}

\section*{Acknowledgment}
We acknowledge funding from the European Research Council (ERC) under the European Union's Horizon 2020 research and innovation programme (agreement No. 851720). 

\balance

\bibliographystyle{IEEEtranS}
\bibliography{main}

\begin{thebibliography}{10}
\providecommand{\url}[1]{#1}
\csname url@samestyle\endcsname
\providecommand{\newblock}{\relax}
\providecommand{\bibinfo}[2]{#2}
\providecommand{\BIBentrySTDinterwordspacing}{\spaceskip=0pt\relax}
\providecommand{\BIBentryALTinterwordstretchfactor}{4}
\providecommand{\BIBentryALTinterwordspacing}{\spaceskip=\fontdimen2\font plus
\BIBentryALTinterwordstretchfactor\fontdimen3\font minus
  \fontdimen4\font\relax}
\providecommand{\BIBforeignlanguage}[2]{{%
\expandafter\ifx\csname l@#1\endcsname\relax
\typeout{** WARNING: IEEEtranS.bst: No hyphenation pattern has been}%
\typeout{** loaded for the language `#1'. Using the pattern for}%
\typeout{** the default language instead.}%
\else
\language=\csname l@#1\endcsname
\fi
#2}}
\providecommand{\BIBdecl}{\relax}
\BIBdecl

\bibitem{chatgpt}
``Chatgpt,'' \url{https://openai.com/blog/chatgpt}, accessed: 2023-03-27.

\bibitem{copilot}
``Copilot website,'' \url{https://copilot.github.com}, accessed: 2022-11-10.

\bibitem{prettier}
``Prettier,'' \url{https://prettier.io}, accessed: 2023-03-25.

\bibitem{replication}
``\url{https://github.com/CodeReviewAutomationSota/code_review_automation_sota}.''

\bibitem{al2020workload}
W.~H.~A. Al{-}Zubaidi, P.~Thongtanunam, H.~K. Dam, C.~Tantithamthavorn, and
  A.~Ghose, ``Workload-aware reviewer recommendation using a multi-objective
  search-based approach,'' in \emph{16th {ACM} International Conference on
  Predictive Models and Data Analytics in Software Engineering, {PROMISE}},
  2020, pp. 21--30.

\bibitem{alon2019code2vec}
U.~Alon, M.~Zilberstein, O.~Levy, and E.~Yahav, ``Code2vec: Learning
  distributed representations of code,'' \emph{Proc. ACM Program. Lang.},
  vol.~3, no. POPL, pp. 40:1--40:29, 2019.

\bibitem{asthana:esecfse2019}
S.~Asthana, R.~Kumar, R.~Bhagwan, C.~Bird, C.~Bansal, C.~S. Maddila, S.~Mehta,
  and B.~Ashok, ``Whodo: Automating reviewer suggestions at scale,'' in
  \emph{27th {ACM} Joint Meeting on European Software Engineering Conference
  and Symposium on the Foundations of Software Engineering, {ESEC/FSE}}, 2019,
  p. 937–945.

\bibitem{bacchelli2013expectations}
A.~Bacchelli and C.~Bird, ``Expectations, outcomes, and challenges of modern
  code review,'' in \emph{Proceedings of the 2013 international conference on
  software engineering}.\hskip 1em plus 0.5em minus 0.4em\relax IEEE Press,
  2013, pp. 712--721.

\bibitem{Bavota:icsme2015}
G.~Bavota and B.~Russo, ``Four eyes are better than two: On the impact of code
  reviews on software quality,'' in \emph{31th {IEEE} International Conference
  on Software Maintenance and Evolution, {ICSME}}, 2015, pp. 81--90.

\bibitem{belleretal:msr2014}
M.~Beller, A.~Bacchelli, A.~Zaidman, and E.~Juergens, ``Modern code reviews in
  open-source projects: Which problems do they fix?'' in \emph{11th {IEEE/ACM}
  Working Conference on Mining Software Repositories, {MSR}}, 2014, pp.
  202--211.

\bibitem{Bosu:2013}
A.~Bosu and J.~C. Carver, ``Impact of peer code review on peer impression
  formation: A survey,'' in \emph{7th {IEEE/ACM} International Symposium on
  Empirical Software Engineering and Measurement, {ESEM}}, 2013, pp. 133--142.

\bibitem{chouchen:asc2021}
M.~Chouchen, A.~Ouni, M.~W. Mkaouer, R.~G. Kula, and K.~Inoue, ``Whoreview: A
  multi-objective search-based approach for code reviewers recommendation in
  modern code review,'' \emph{Applied Soft Computing}, vol. 100, p. 106908,
  2021.

\bibitem{coxon1999sorting}
A.~Coxon, \emph{Sorting Data: Collection and Analysis}, ser. Quantitative
  Applications in the Social Sciences, 1999, no. no. 127.

\bibitem{Falleri:ase2014}
J.~Falleri, F.~Morandat, X.~Blanc, M.~Martinez, and M.~Monperrus,
  ``Fine-grained and accurate source code differencing,'' in \emph{29th
  {IEEE/ACM} International Conference on Automated Software Engineering,
  {ASE}}, 2014, pp. 313--324.

\bibitem{Fraser:fse2011}
G.~Fraser and A.~Arcuri, ``Evosuite: automatic test suite generation for
  object-oriented software,'' in \emph{21st {ACM} Joint Meeting of the European
  Software Engineering Conference and the {ACM/SIGSOFT} Symposium on the
  Foundations of Software Engineering, {ESEC-FSE}}, 2011, pp. 416--419.

\bibitem{grissom:lawrence2005}
R.~J. Grissom and J.~J. Kim, \emph{Effect sizes for research: A broad practical
  approach.}, 2005.

\bibitem{holm:sjs1979}
S.~Holm, ``A simple sequentially rejective bonferroni test procedure,''
  \emph{Scandinavian Journal on Statistics}, vol.~6, no.~2, pp. 65--70, 1979.

\bibitem{hong:fse2022}
Y.~Hong, C.~Tantithamthavorn, P.~Thongtanunam, and A.~Aleti, ``Commentfinder: A
  simpler, faster, more accurate code review comments recommendation,'' in
  \emph{30th {ACM} Joint European Software Engineering Conference and Symposium
  on the Foundations of Software Engineering, {ESEC-FSE}}, 2022, p. 507–519.

\bibitem{jiang:jss2019}
J.~Jiang, D.~Lo, J.~Zheng, X.~Xia, Y.~Yang, and L.~Zhang, ``Who should make
  decision on this pull request? analyzing time-decaying relationships and file
  similarities for integrator prediction,'' \emph{J. Syst. Softw.}, vol. 154,
  no.~C, p. 196–210, 2019.

\bibitem{jiang:ist2017}
\BIBentryALTinterwordspacing
J.~Jiang, Y.~Yang, J.~He, X.~Blanc, and L.~Zhang, ``Who should comment on this
  pull request? analyzing attributes for more accurate commenter recommendation
  in pull-based development,'' \emph{Inf. Softw. Technol.}, vol.~84, no.~C, p.
  48–62, apr 2017. [Online]. Available:
  \url{https://doi.org/10.1016/j.infsof.2016.10.006}
\BIBentrySTDinterwordspacing

\bibitem{Kudo:sentencePiece}
T.~Kudo and J.~Richardson, ``Sentencepiece: {A} simple and language independent
  subword tokenizer and detokenizer for neural text processing,'' in \emph{8th
  Conference on Empirical Methods in Natural Language Processing, {EMNLP}},
  2018, pp. 66--71.

\bibitem{li_lingwei:esecfse2022}
L.~Li, L.~Yang, H.~Jiang, J.~Yan, T.~Luo, Z.~Hua, G.~Liang, and C.~Zuo,
  ``Auger: Automatically generating review comments with pre-training models,''
  in \emph{30th {ACM} Joint European Software Engineering Conference and
  Symposium on the Foundations of Software Engineering, {ESEC/FSE}}, 2022, p.
  1009–1021.

\bibitem{li:esecfse2022}
Z.~Li, S.~Lu, D.~Guo, N.~Duan, S.~Jannu, G.~Jenks, D.~Majumder, J.~Green,
  A.~Svyatkovskiy, S.~Fu, and N.~Sundaresan, ``Automating code review
  activities by large-scale pre-training,'' in \emph{30th {ACM} Joint European
  Software Engineering Conference and Symposium on the Foundations of Software
  Engineering, {ESEC/FSE}}, 2022, pp. 1035--1047.

\bibitem{mantyla:tse2009}
M.~V. M\"antyl\"a and C.~Lassenius, ``What types of defects are really
  discovered in code reviews?'' \emph{IEEE Transactions on Software
  Engineering}, vol.~35, no.~3, pp. 430--448, 2009.

\bibitem{mastropaolo:icse2022}
A.~Mastropaolo, L.~Pascarella, and G.~Bavota, ``Using deep learning to generate
  complete log statements,'' in \emph{44th {IEEE/ACM} International Conference
  on Software Engineering, {ICSE}}, 2022, pp. 2279--2290.

\bibitem{McIntosh:msr2014}
S.~McIntosh, Y.~Kamei, B.~Adams, and A.~E. Hassan, ``The impact of code review
  coverage and code review participation on software quality: A case study of
  the qt, vtk, and itk projects,'' in \emph{11th {IEEE/ACM} Working Conference
  on Mining Software Repositories, {MSR}}, 2014, pp. 192--201.

\bibitem{mirsaeedi:icse2020}
E.~Mirsaeedi and P.~C. Rigby, ``Mitigating turnover with code review
  recommendation: Balancing expertise, workload, and knowledge distribution,''
  in \emph{42nd {ACM/IEEE} International Conference on Software Engineering,
  {ICSE}}, 2020, p. 1183–1195.

\bibitem{morales2015saner}
R.~Morales, S.~McIntosh, and F.~Khomh, ``Do code review practices impact design
  quality? a case study of the qt, vtk, and itk projects,'' in \emph{Proc. of
  the 22nd Int'l Conf. on Software Analysis, Evolution, and Reengineering
  (SANER)}, 2015, pp. 171--180.

\bibitem{ouni:icsme2016}
A.~Ouni, R.~G. Kula, and K.~Inoue, ``Search-based peer reviewers recommendation
  in modern code review,'' in \emph{32nd {IEEE} International Conference on
  Software Maintenance and Evolution, {ICSME}}, 2016, pp. 367--377.

\bibitem{PachecoE2007Poster}
C.~Pacheco and M.~D. Ernst, ``{Randoop:} feedback-directed random testing for
  {Java},'' in \emph{{ACM/SIGPLAN} International Symposium on New Ideas, New
  Paradigms, and Reflections on Programming and Software, {OOPSLA}}, 2007, pp.
  815--816.

\bibitem{papineni:acl2002}
K.~Papineni, S.~Roukos, T.~Ward, and W.-J. Zhu, ``Bleu: A method for automatic
  evaluation of machine translation,'' in \emph{40th Annual Meeting on
  Association for Computational Linguistics, {ACL}}, 2002, pp. 311--318.

\bibitem{Pascarella:CSCW2018}
L.~Pascarella, D.~Spadini, F.~Palomba, M.~Bruntink, and A.~Bacchelli,
  ``Information needs in contemporary code review,'' vol.~2, no. CSCW, 2018.

\bibitem{raffel:jmlr2019}
C.~Raffel, N.~Shazeer, A.~Roberts, K.~Lee, S.~Narang, M.~Matena, Y.~Zhou,
  W.~Li, and P.~J. Liu, ``Exploring the limits of transfer learning with a
  unified text-to-text transformer,'' \emph{J. Mach. Learn. Res.}, vol.~21, pp.
  140:1--140:67, 2020.

\bibitem{rahman:icsec2016}
M.~M. Rahman, C.~K. Roy, and J.~A. Collins, ``Correct: code reviewer
  recommendation in github based on cross-project and technology experience,''
  in \emph{38th International Conference on Software Engineering, {ICSE}},
  2016, pp. 222--231.

\bibitem{Rigby:fse2013}
P.~C. Rigby and C.~Bird, ``Convergent contemporary software peer review
  practices,'' in \emph{21st {ACM/SIGSOFT} Joint Meeting of the European
  Software Engineering Conference and the Symposium on the Foundations of
  Software Engineering, {ESEC-FSE}}, 2013, pp. 202--212.

\bibitem{Rigby:tosem2014}
P.~C. Rigby, D.~M. Germ{\'{a}}n, L.~L.~E. Cowen, and M.~D. Storey, ``Peer
  review on open-source software projects: Parameters, statistical models, and
  theory,'' \emph{ACM Trans. Softw. Eng. Methodol.}, vol.~23, no.~4, pp.
  35:1--35:33, 2014.

\bibitem{Rosner2011}
B.~Rosner, \emph{Fundamentals of Biostatistics}, 7th~ed.\hskip 1em plus 0.5em
  minus 0.4em\relax Brooks/Cole, Boston, MA, 2011.

\bibitem{Caitlin:icse2018}
C.~Sadowski, E.~S{\"{o}}derberg, L.~Church, M.~Sipko, and A.~Bacchelli,
  ``Modern code review: a case study at google,'' in \emph{40th International
  Conference on Software Engineering: Software Engineering in Practice, {ICSE}
  {(SEIP)}}, 2018, pp. 181--190.

\bibitem{shu:aaai2019}
S.~Shi, M.~Li, D.~Lo, F.~Thung, and X.~Huo, ``Automatic code review by learning
  the revision of source code,'' in \emph{The Thirty-Third {AAAI} Conference on
  Artificial Intelligence, {AAAI} 2019}, 2019, pp. 4910--4917.

\bibitem{strand:icseseip2020}
A.~Strand, M.~Gunnarson, R.~Britto, and M.~Usman, ``Using a context-aware
  approach to recommend code reviewers: findings from an industrial case
  study,'' in \emph{42nd International Conference on Software Engineering,
  Software Engineering in Practice, {ICSE-SEIP}}, 2020, pp. 1--10.

\bibitem{thongtanunam:icse2022}
P.~Thongtanunam, C.~Pornprasit, and C.~Tantithamthavorn, ``Autotransform:
  Automated code transformation to support modern code review process,'' in
  \emph{2022 IEEE/ACM 44th International Conference on Software Engineering
  (ICSE)}, 2022, pp. 237--248.

\bibitem{thongtanunam:saner2015}
P.~Thongtanunam, C.~Tantithamthavorn, R.~G. Kula, N.~Yoshida, H.~Iida, and
  K.~Matsumoto, ``Who should review my code? {A} file location-based
  code-reviewer recommendation approach for modern code review,'' in \emph{22nd
  {IEEE} International Conference on Software Analysis, Evolution, and
  Reengineering, {SANER}}, 2015, pp. 141--150.

\bibitem{tian:icsme2022}
F.~Tian and C.~Treude, ``Adding context to source code representations for deep
  learning,'' in \emph{{IEEE} International Conference on Software Maintenance
  and Evolution, {ICSME}}, 2022, pp. 374--378.

\bibitem{Tsantalis:tse2022}
N.~Tsantalis, A.~Ketkar, and D.~Dig, ``Refactoringminer 2.0,'' \emph{{IEEE}
  Trans. Software Eng.}, vol.~48, no.~3, pp. 930--950, 2022.

\bibitem{tufano:icse2019}
M.~Tufano, J.~Pantiuchina, C.~Watson, G.~Bavota, and D.~Poshyvanyk, ``On
  learning meaningful code changes via neural machine translation,'' in
  \emph{41st {IEEE/ACM} International Conference on Software Engineering,
  {ICSE}}, 2019, pp. 25--36.

\bibitem{tufano:tosem2019}
M.~Tufano, C.~Watson, G.~Bavota, M.~{Di Penta}, M.~White, and D.~Poshyvanyk,
  ``An empirical study on learning bug-fixing patches in the wild via neural
  machine translation,'' \emph{{ACM} Trans. Softw. Eng. Methodol.}, vol.~28,
  no.~4, pp. 19:1--19:29, 2019.

\bibitem{tufano:icse2022}
R.~Tufano, S.~Masiero, A.~Mastropaolo, L.~Pascarella, D.~Poshyvanyk, and
  G.~Bavota, ``Using pre-trained models to boost code review automation,'' in
  \emph{44th {IEEE/ACM} International Conference on Software Engineering,
  {ICSE}}, 2022, pp. 2291--2302.

\bibitem{tufano:icse2021}
R.~Tufano, L.~Pascarella, M.~Tufano, D.~Poshyvanyk, and G.~Bavota, ``Towards
  automating code review activities,'' in \emph{43rd {IEEE/ACM} International
  Conference on Software Engineering, {ICSE}}, 2021, pp. 163--174.

\bibitem{wilcoxon:ibs1992}
F.~Wilcoxon, ``Individual comparisons by ranking methods,'' \emph{International
  Biometric Society, Wiley}, vol.~1, no.~6, pp. 80--83, 1945.

\bibitem{xia:icsm2015}
X.~Xia, D.~Lo, X.~Wang, and X.~Yang, ``Who should review this change?: Putting
  text and file location analyses together for more accurate recommendations,''
  in \emph{31th {IEEE} International Conference on Software Maintenance and
  Evolution, {ICSME}}, 2015, pp. 261--270.

\bibitem{xia:iwsm2017}
Z.~Xia, H.~Sun, J.~Jiang, X.~Wang, and X.~Liu, ``A hybrid approach to code
  reviewer recommendation with collaborative filtering,'' in \emph{6th
  International Workshop on Software Mining, {SoftwareMining}}, 2017, pp.
  24--31.

\bibitem{ying:csise2016}
H.~Ying, L.~Chen, T.~Liang, and J.~Wu, ``Earec: leveraging expertise and
  authority for pull-request reviewer recommendation in github,'' in \emph{3rd
  International Workshop on CrowdSourcing in Software Engineering,
  {CSI-SE}@{ICSE}}, 2016, pp. 29--35.

\bibitem{Shiwen:jss2022}
S.~Yu, T.~Wang, and J.~Wang, ``Data augmentation by program transformation,''
  \emph{J. Syst. Softw.}, vol. 190, p. 111304, 2022.

\bibitem{yu:ist2016}
Y.~Yu, H.~Wang, G.~Yin, and T.~Wang, ``Reviewer recommendation for
  pull-requests in github: What can we learn from code review and bug
  assignment?'' \emph{Inf. Softw. Technol.}, vol.~74, pp. 204--218, 2016.

\bibitem{zanjanitse:2016}
M.~B. Zanjani, H.~Kagdi, and C.~Bird, ``Automatically recommending peer
  reviewers in modern code review,'' \emph{IEEE Transactions on Software
  Engineering}, vol.~42, no.~6, pp. 530--543, 2016.

\end{thebibliography}
\newpage

\begin{IEEEbiography}[{\includegraphics[width=1in,height=1.25in,clip,keepaspectratio]{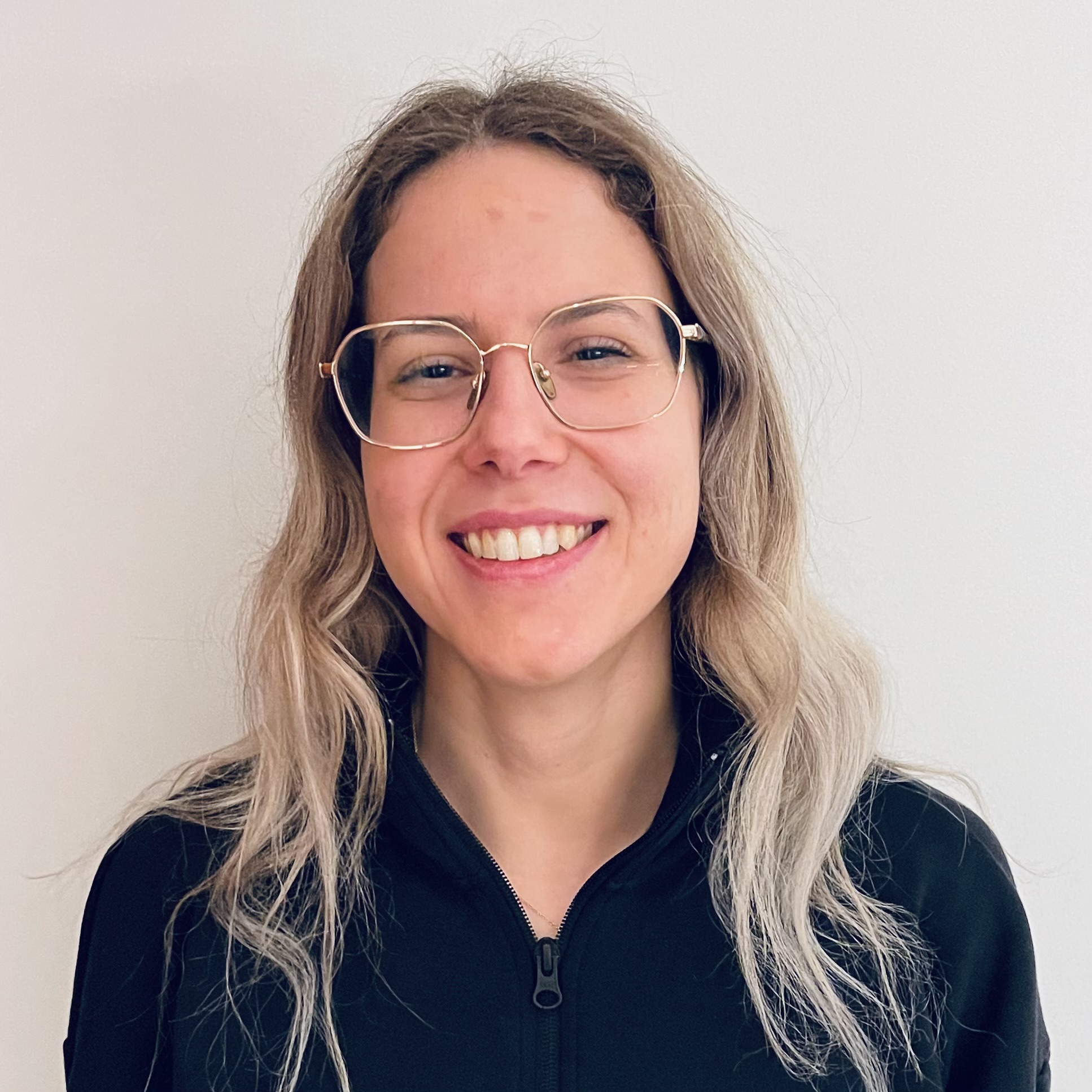}}]{Rosalia Tufano} is a Ph.D student in the Faculty of Informatics at the Università della Svizzera Italiana (USI), Switzerland, and part of the Software Analytics Research Team (SEART). She received her MSc. in Applied Mathematics from Università degli Studi di Napoli Federico II , Italy, in March 2019. Her research interests mainly include the study and the application of machine learning techniques to support code-related tasks. More information available at: \url{https://www.inf.usi.ch/phd/tufanr/}
\end{IEEEbiography}

\begin{IEEEbiography}[{\includegraphics[width=1in,height=1.25in,clip,keepaspectratio]{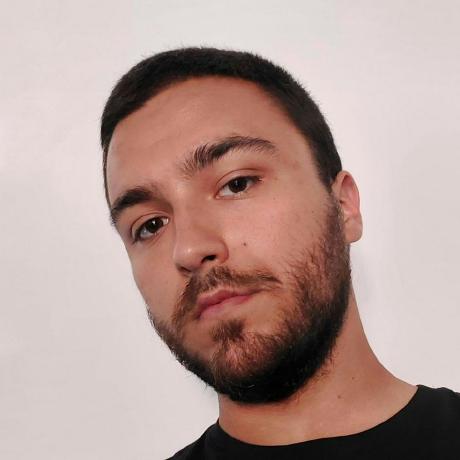}}]{Ozren Dabi\'c} is a Research Assistant and Software Engineer for the Software Analytics Research Team (SEART) of the Software Institute in Lugano. As a Full Stack Developer, his work primarily revolves around creating software solutions and web platforms for academia, intended to streamline the research process. He also participates in research focusing on the usage of Deep Learning models to automate various software engineering tasks.
\end{IEEEbiography}

\begin{IEEEbiography}[{\includegraphics[width=1in,height=1.25in,clip,keepaspectratio]{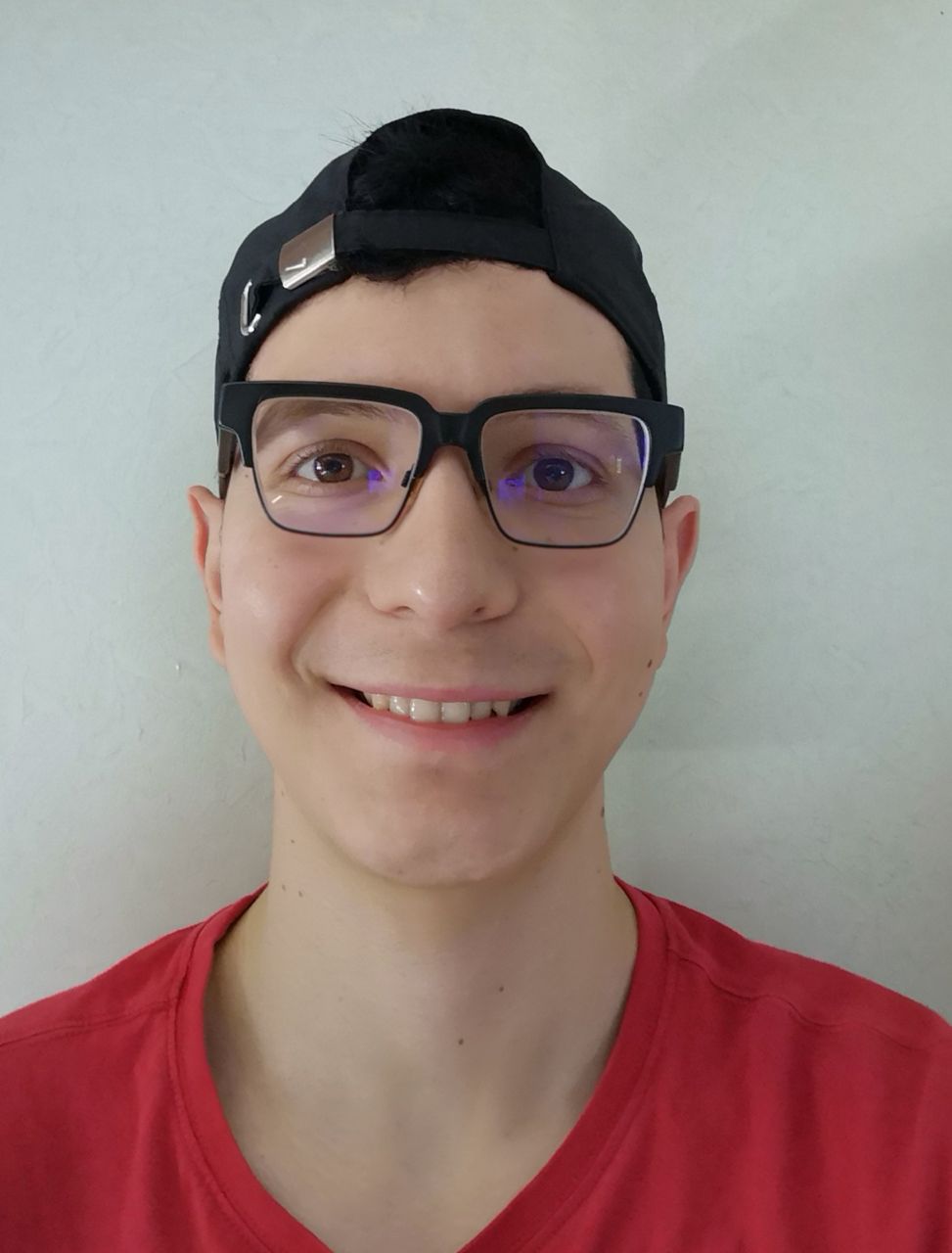}}]{Antonio Mastropaolo} is a Ph.D. student in the Faculty of Informatics at the Universit\`a della Svizzera italiana (USI), Switzerland, where he is part of the Software Institute. He received his MSc. in Software System Security from Universit\`a degli studi del Molise, Italy, in July 2020. His research interests include the study and the application of deep-learning techniques to foster code-related tasks. More information available at: \url{https://antoniomastropaolo.com}.
\end{IEEEbiography}

\begin{IEEEbiography}[{\includegraphics[width=1in,height=1.25in,clip,keepaspectratio]{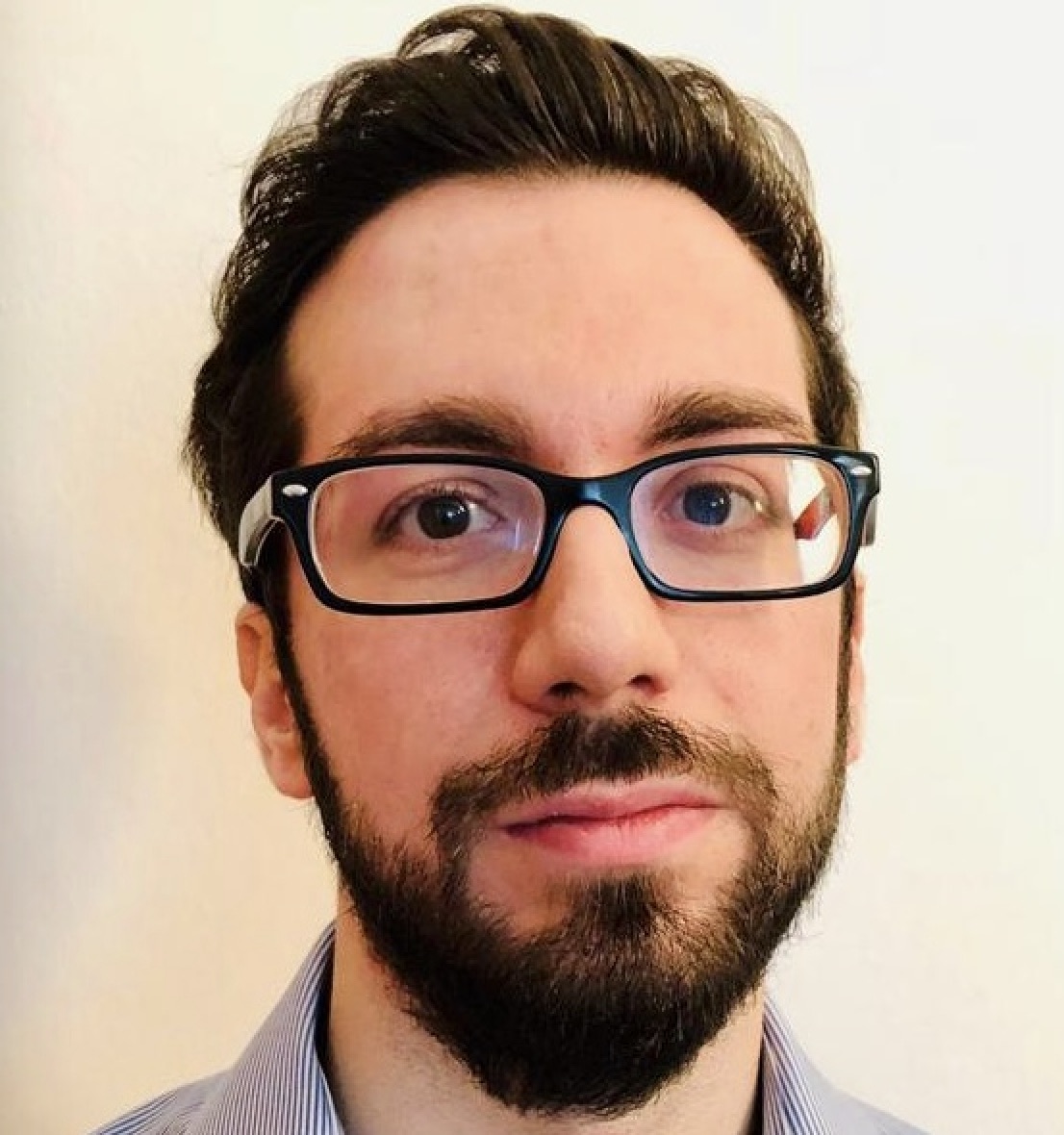}}]{Matteo Ciniselli} is a Ph.D. student in the Faculty of Informatics at the Università della Svizzera Italiana (USI), Switzerland. He received his MSc. in Mathematical Engineering from Politecnico di Milano, Italy, in April 2015. His research interests mainly include the study of deep-learning models to improve the performances of code-related tasks. More information available at: \url{https://www.inf.usi.ch/phd/cinism/}
\end{IEEEbiography}

\begin{IEEEbiography}[{\includegraphics[width=1in,height=1.25in,clip,keepaspectratio]{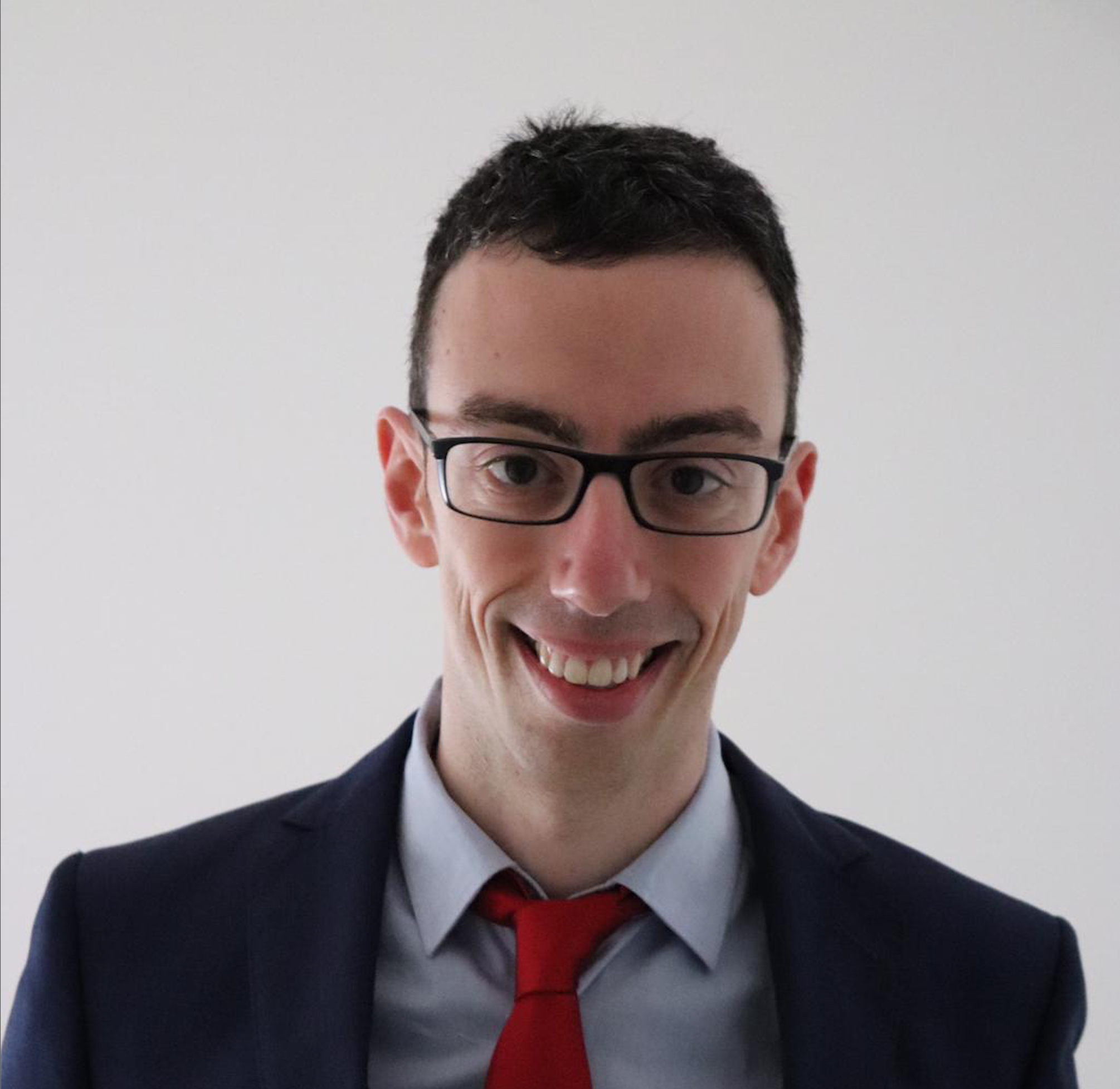}}]{Gabriele Bavota} is an associate professor at the Faculty of Informatics of the Universit\`a della Svizzera italiana (USI), Switzerland, where he is part of the Software Institute and he leads the SEART research group. He received the PhD in Computer Science from the University of Salerno, Italy, in 2013. His research interests include software maintenance and evolution, code quality, mining software repositories, and empirical software engineering. On these topics, he authored over 150 papers appeared in international journals and conferences and has received four ACM Sigsoft Distinguished Paper awards at the three top software engineering conferences: ASE 2013 and 2017, ESEC-FSE 2015, and ICSE 2015. He also received the best/distinguished paper award at SCAM 2012, ICSME 2018, MSR 2019, and ICPC 2020.
He is the recipient of the 2018 ACM Sigsoft Early Career Researcher Award for outstanding contributions in the area of software engineering as an early career investigator and the principal investigator of the DEVINTA ERC project. More information is available at: \url{https://www.inf.usi.ch/faculty/bavota/}.
\end{IEEEbiography}

\end{document}